\documentclass[prb, aps, preprint, showpacs]{revtex4}

\usepackage{graphicx}
\usepackage{subfigure}
\usepackage{bm}
\usepackage{amssymb}
\usepackage{amsmath}

\begin{document}

\title{Nature of Stripes in the Generalized $t-J$ Model Applied to The
Cuprate Superconductors}
\author{Kai-Yu Yang$^{1,2}$, Wei Qiang Chen $^{2}$, T. M. Rice $^{1,2}$, M.
Sigrist $^{1}$, Fu-Chun Zhang $^{2}$} \affiliation{$^{1}$ Institut
f$\ddot{u}$r Theoretische Physik, ETH Z$\ddot{u}$rich,
CH-8093 Z$\ddot{u}$rich, Switzerland \\
$^{2}$ Center for Theoretical and Computational Physics and Department of
Physics, The University of Hong Kong, Hong Kong SAR, China }

\begin{abstract}
Recent transport properties on the stripe phase in
La$_{\text{1.875}}$Ba$_{\text{01.25}}$CuO$_{\text{4}}$ by Li
\textit{et al.} \cite{Li-PRL-07} found 2-dimensional
superconductivity over a wide temperature range including a
Berezinski-Kosterlitz-Thouless transition at a temperature T=16K,
with 3-dimensional superconducting (SC) ordering only at T=4K.
These results contradict the long standing belief that the onset
of superconductivity is suppressed by stripe ordering and suggest
coexistence of stripe and SC phases. The lack of 3-D
superconducting order above T=4K requires an antiphase ordering in
the SC state to suppress the interlayer Josephson coupling as
proposed by Berg \textit{et al.} \cite{Berg-PRL-07}. Here we use a
renormalized mean field theory for a generalized t-J model to
examine in detail the energetics of the spin and charge stripe
ordered SC states including possible antiphase domains in the SC
order. We find that the energies of these modulated states are
very close to each other and that the anisotropy present in the
low temperature tetragonal crystal structure favors stripe
resonating valence bond states.  The stripe antiphase SC states
are found to have energies very close,but always above, the ground
state energy which suggests additional physical effects are
responsible for their stability.

\end{abstract}

\date{today}
\maketitle

\section{Introduction}

Recently Li and coworkers \cite{Li-PRL-07} reported new results on transport
properties of the stripe phase in La$_{\text{1.875}}$Ba$_{\text{0.125}}$CuO$%
_{\text{4}}$. They found that 2-dimensional superconducting (SC)
fluctuations appear at an onset temperature
T$_{\text{c}}^{\text{2D}}$(=42K) which greatly exceeds the
critical temperature for 3-dimensional SC order, T$_{\text{c}} $
(=4K). These results contradicted the long standing belief that
the onset of SC behavior was suppressed to very low temperatures
in the presence of the static spin and charge density wave (SDW
and CDW hereafter) or more precisely spin and charge stripe
orderings. Li \textit{et al.,} \cite{Li-PRL-07} found strong
evidence for a Berezinskii-Kosterlitz-Thouless transition (BKT) at
T$_{\text{BKT}}$ (=16K). This implies that the Josephson coupling
between the CuO$_{\text{2}}$ planes strictly vanishes for
T$>$T$_{\text{c}}$. Shortly afterwards Berg \textit{et al.}
\cite{Berg-PRL-07} proposed that the strict interplanar decoupling
arises because the planar superconductivity contains a periodic
array of lines of $\pi $-phase shift which rotate through $\pi /2$
up the c-axis together with the spin and charge stripe ordering in
the low temperature tetragonal (LTT) phase. SDW order also appears
at the same onset temperature, $T_{c}^{2D}$ in zero magnetic field
and this temperature is clearly separated from the crystallographic
transition temperature $T_{co}$ separating the low temperature
orthorhombic (LTO) and LTT phases. In this material the LTT phase
shows a superlattice ordering at all temperatures below $T_{co}
$. \cite{Kim-PRB-08} Note however recent experiments by Fink \textit{et al.} on La$_{\text{%
1.8-x}}$Eu$_{\text{0.2}}$Sr$_{\text{x}}$CuO$_{4}$
\cite{Fink-arXiv-08} found different temperatures with the
superlattice onset below the crystallographic phase transition
temperature. Earlier studies by Lee \textit{et al.} on
superoxygenated La$_{\text{2}}$CuO$_{\text{4}}$
\cite{YSLee-PRB-99} found the same onset temperature for both SC
and SDW order (T=42K). They also noted that signs of a CDW
superlattice at higher temperature (T=55K) has been reported.
These temperatures coincide with
the values found by Li \textit{et al.} in La$_{\text{1.875}}$Ba$_{\text{0.125%
}}$CuO$_{\text{4}}$ which suggests that Lee \textit{et al.} were observing
a similar stripe order with coexisting SDW and SC. In this case, however, the SC order is 3-dimensional, consistent with the absence of $\pi/2$-rotations in the crystal
structure. These experiments lead us to conclude that in the presence of a CDW superlattice, coexisting SDW and antiphase d-wave SC can be favored.

Actually a similar ordering was suggested on general grounds earlier by
Zhang \cite{Zhang-JPCS-98} and also by Himeda , Kato and Ogata \cite{Himeda-PRL-02} on the
basis of variational Monte Carlo calculations (VMC) for the strongly
correlated one band $t-t^{\prime }-J$ model. Himeda \textit{et al} \cite%
{Himeda-PRL-02} found that a modulated state with combined SDW,
and CDW and d-wave superconductivity (dSC) containing site- or bond- centered anti-phase
domain walls ($\pi$DW) ( a state we denote as SDW+CDW+APdSC$^{s/b}$) had a
lower energy than a uniform d-wave SC state over a wide range of
parameters and was even lower than a modulated state without
anti-phase (denoted as SDW+CDW+dSC$^{s/b}$) in a narrower parameter range.
Recent VMC and renormalized mean field theory (RMFT) calculations
\cite{Raczkowski-PRB-07} have found that CDW+APdSC$^{s/b}$ state ($\pi
$-DRVB state in ref.\cite{Raczkowski-PRB-07}) cost surprisingly
little energy even in the absence of SDW modulations.

In this paper we report on calculations using the RMFT method to examine in
greater detail the energetics of these novel modulated states within the
generalized $t-t^{\prime }-t^{\prime \prime }-J$ model. This method
approximates the strong correlation condition of no double occupancy by
Gutzwiller renormalization factors and generally agrees well with full VMC
calculations which treat the strong correlation condition exactly. The
static stripe phase appears in the LTT phase
of La$_{\text{1.875}}$Ba$_{\text{0.125}}$CuO$_{\text{4}}$. This
crystallographic phase is entered at a temperature T$_{\text{co}}$ (=52K $>$%
T $_{\text{c}}^{\text{2D}}$) and displays a complex crystal structure which
has not been fully determined to the best of our knowledge. Note that
although the overall crystal structure is tetragonal the individual CuO$_{%
\text{2}}$ planes do not have square symmetry. Along one (x-) axis
the Cu-O-Cu bonds are straight but in the perpendicular direction
they are buckled \cite{Buchner-PRL-94}. Since the Cu-Cu distance
is required to be the same in both directions there is a
compressive stress along the x-axis which may well be the origin
of the CDW superlattice that appears at the crystallographic phase
transition into the LTT phase. At present the detailed
displacements inside the supercell have not been refined. In our
calculations we introduce a site dependent potential shift to
mimic this effect. In addition we examine the effect of the
hopping anisotropy between x- and y-axes which results from the
different Cu-O-Cu bonding in the x and y directions. Such
anisotropy was also considered by Capello \textit{et al.}
\cite{Capello-PRB-08} in their work on stripes made from
anti-phase shifts in the superconductivity.

\section{Renormalized Mean Field Theory for the extended $t - J$ model}

The $t-J$ model was introduced in the early days of cuprate research by
Anderson and by Zhang and Rice to describe lightly hole doped CuO$_{\text{2}%
} $ planes \cite{t-J-model}. In this single band model configurations with doubly occupied
sites are strictly forbidden due to the strong onsite Coulomb repulsion. The
Hamiltonian takes the form, suppressing the constraint

\begin{eqnarray}
H_{tj}&=&-\sum_{\left( i,j\right) ,\sigma }t_{\left( i,j\right) }\left( \hat{c}%
_{i,\sigma }^{\dag }\hat{c}_{j,\sigma }+h.c.\right) +\sum_{\langle
i,j\rangle }J_{\langle i,j\rangle }\mathbf{\hat{S}}_{i}\cdot \mathbf{\hat{S}}
_{j} \notag \\ &&+\sum_{i}V_{i}\hat{n}_{i}.  \label{eq:tj}
\end{eqnarray}%
In the first term we include hopping processes between nearest
neighboring (nn) sites (denoted by $\langle i,j\rangle $), next
neighboring sites (nnn) and 3rd neighboring sites (nnnn) on a
square lattice with matrix elements $t$ , $t^{\prime }$ ,
$t^{\prime \prime }$\ respectively. We will measure all energies
in unit of $t_{0}$ (300 meV) --- a standard value for the nn
hopping matrix element $t$. The superexchange spin-spin
interaction between nn sites $J=0.3$, and $\sigma $ the spin index
takes the value $\pm $. In addition we introduce a potential shift
$V_{i}$ which varies from site to site within the supercell to
mimic the effect of the crystallographic superlattice in the LTT
crystal structure. The strong coupling constraint of no double
occupancy is very difficult to treat analytically. Zhang and
coworkers introduced Gutzwiller renormalization factors to
approximate the constraint \cite{RMFT}. This approximation has
been shown to be quite accurate for mean field theories when
compared to numerical evaluations by VMC
of expectation values of the corresponding mean field wavefunctions, $%
\left\vert \Psi \right\rangle $, which are exactly projected down to the
constrained Hilbert space \cite{RMFT-VMC}. Later the case of AF ordering was considered by
Himeda and Ogata, who showed that an anisotropic spin renormalization term
is required to reproduce the VMC results \cite{Himeda-Ogata}. The resulting renormalized
Hamiltonian is
\begin{eqnarray}
H &=&-\sum_{\left( i,j\right) ,\sigma }g_{\left( i,j\right) ,\sigma
}^{t}t_{\left( i,j\right) }\left( \hat{c}_{i,\sigma }^{\dag }\hat{c}%
_{j,\sigma }+h.c.\right) \notag
\\
&&+\sum_{\langle i,j\rangle }J_{\langle i,j\rangle }\left[ g_{\langle
i,j\rangle }^{s,z}\hat{S}_{i}^{z}\hat{S}_{j}^{z}+g_{\langle i,j\rangle
}^{s,xy}\left( \hat{S}_{i}^{+}\hat{S}_{j}^{-}+\hat{S}_{i}^{-}\hat{S}%
_{j}^{+}\right) /2\right]  \notag
\\ &&   +\sum_{i}V_{i}\hat{n}_{i} .   \label{eq:GZHamiltonian}
\end{eqnarray}%
The renormalization factors $g^{t}$, $g^{s,xy}$ and $g^{s,z}$ used
to evaluate a projected mean field wavefunction depend on the
local values of the magnetic and pairing order parameters and the
local kinetic energy and hole density which are defined as follows
\begin{eqnarray}
m_{i} &=&\left\langle \Psi _{0}\right\vert \hat{S}_{i}^{z}\left\vert \Psi
_{0}\right\rangle ;\notag \\
\Delta _{\left\langle i,j\right\rangle ,\sigma }&=&\sigma
\left\langle \Psi _{0}\right\vert \hat{c}_{i,\sigma
}\hat{c}_{j,-\sigma }\left\vert \Psi _{0}\right\rangle ; \notag\\
\chi _{\left( i,j\right) ,\sigma } &=&\left\langle \Psi _{0}\right\vert \hat{%
c}_{i,\sigma }^{\dag }\hat{c}_{j,\sigma }\left\vert \Psi
_{0}\right\rangle ;\notag \\
 \delta_{i}&=&1-\left\langle \Psi
_{0}\right\vert \hat{n}_{i}\left\vert \Psi _{0}\right\rangle ,
\label{eq:orderP}
\end{eqnarray}
where $\left\vert \Psi _{0}\right\rangle $ is the unprojected
wavefunction. The two pairing amplitudes $\Delta _{\left\langle
i,j\right\rangle ,\sigma =\pm }$ are treated independently to
incorporate a possible triplet component. The explicit
renormalization factors introduced first by Himeda and Ogata are
quite complex, \cite{Himeda-Ogata} and we use here a simpler form
as follows,\cite{simplify-G}

\begin{eqnarray}
g_{\left( i,j\right) ,\sigma }^{t} &=&g_{i,\sigma }^{t}g_{j,\sigma }^{t}; \notag \\
g_{i,\sigma }^{t}&=&\sqrt{\frac{%
2\delta _{i}\left( 1-\delta _{i}\right) }{1-\delta _{i}^{2}+4m_{i}^{2}}\frac{%
1+\delta _{i}+\sigma 2m_{i}}{1+\delta _{i}-\sigma 2m_{i}}};  \notag \\
g_{\langle i,j\rangle }^{s,xy} &=&g_{i}^{s,xy}g_{j}^{s,xy};\notag \\
g_{i}^{s,xy}&=&\frac{2\left( 1-\delta _{i}\right) }{1-\delta
_{i}^{2}+4m_{i}^{2}};  \notag \\
g_{\langle i,j\rangle }^{z} &=&g_{\langle i,j\rangle
}^{s,xy}\frac{2\left( \overline{\Delta}_{\langle i,j\rangle
}^{2}+\overline{\chi}_{\langle i,j\rangle
}^{2}\right) -4m_{i}m_{j}X_{\langle i,j\rangle }^{2}}{2\left( \overline{\Delta}%
_{\langle i,j\rangle }^{2}+\overline{\chi}_{\langle i,j\rangle
}^{2}\right)
-4m_{i}m_{j}};\notag \\
X_{\langle i,j\rangle } &=&1+\frac{12\left( 1-\delta _{i}\right) \left(
1-\delta _{j}\right) \left( \overline{\Delta}_{\langle i,j\rangle }^{2}+\overline{\chi}%
_{\langle i,j\rangle }^{2}\right) }{\sqrt{\left( 1-\delta
_{i}^{2}+4m_{i}^{2}\right) \left( 1-\delta _{j}^{2}+4m_{j}^{2}\right) }} ,\notag \\
\label{eq:gfactors}
\end{eqnarray}
where $\overline{\Delta}_{\langle i,j\rangle }=\sum_{\sigma
}\Delta _{\left\langle i,j\right\rangle ,\sigma }/2$,
$\overline{\chi}_{\langle i,j\rangle }=\sum_{\sigma }\chi
_{\left\langle i,j\right\rangle ,\sigma }/2$. Since the g-factors
depends on the order parameters, the approach by direct
diagonalization of the mean field Hartree-Fock Hamiltonian
obtained from the Hamiltonian Eq[\ref{eq:GZHamiltonian}] will not
give the best energy of the Hamiltonian
\begin{eqnarray}
&E_{t}& =\left\langle \Psi _{0}\right\vert H\left\vert \Psi
_{0}\right\rangle \notag \\
&=&-\sum_{\left( i,j\right) ,\sigma }g_{\left( i,j\right)
,\sigma }^{t}t_{\left( i,j\right) }\left[ \chi _{\left( i,j\right) ,\sigma
}+h.c.\right]  \notag \\
&&-\sum_{\langle i,j\rangle ,\sigma }J_{\langle i,j\rangle } \left(
\frac{g_{\langle i,j\rangle }^{s,z}}{4}+\frac{g_{\langle i,j\rangle }^{s,xy}%
}{2}\frac{\Delta _{\langle i,j\rangle ,\overline{\sigma}}^{\ast
}}{\Delta _{\langle i,j\rangle ,\sigma }^{\ast }}\right) \Delta
_{\langle i,j\rangle
,\sigma }^{\ast }\Delta _{\langle i,j\rangle ,\sigma } \notag \\
&&-\sum_{\langle i,j\rangle ,\sigma }J_{\langle i,j\rangle }\left( \frac{%
g_{\langle i,j\rangle }^{s,z}}{4}+\frac{g_{\langle i,j\rangle }^{s,xy}}{2}%
\frac{\chi _{\langle i,j\rangle ,\overline{\sigma}}^{\ast }}{\chi
_{\langle i,j\rangle ,\sigma }^{\ast }}\right) \chi _{\langle
i,j\rangle ,\sigma
}^{\ast }\chi _{\langle i,j\rangle ,\sigma } \notag \\
&&+\sum_{i}V_{i}n_{i}+\sum_{\langle i,j\rangle }g_{\langle
i,j\rangle }^{s,z}J_{\langle i,j\rangle }m_{i}m_{j}  \label{eq:energy}
\end{eqnarray}%
Instead, we minimize the energy with respect to the unprojected
wave function $\left\vert \Psi _{0}
\right\rangle $ under the constraints $\sum_{i}{n}_{i}=N_{e}$,
$\left\langle \Psi _{0}|\Psi _{0}\right\rangle =1$, $N_{e}$ is the
total electron density.
That is equivalent to minimizing the
function
\begin{equation}
W=\left\langle \Psi _{0}\right\vert H\left\vert \Psi _{0}\right\rangle
-\lambda \left( \left\langle \Psi _{0}|\Psi _{0}\right\rangle -1\right) -\mu
\left( \sum_{i}\hat{n}_{i}-N_{e}\right)   \label{eq:freeenergy}
\end{equation}%
which results in the following variational relation
\begin{eqnarray}
0&=&\frac{\delta W}{\delta \left\langle \Psi _{0}\right\vert }\notag \\
&=&\sum_{\left(
i,j\right) ,\sigma }\frac{\partial W}{\partial \chi _{\left( i,j\right)
,\sigma }}\frac{\delta \chi _{\left( i,j\right) ,\sigma }}{\delta
\left\langle \Psi _{0}\right\vert }+h.c. \notag \\
&&+\sum_{\left\langle i,j\right\rangle ,\sigma }\frac{\partial
W}{\partial \Delta _{\left\langle i,j\right\rangle
,\sigma }}\frac{\delta \Delta _{\left\langle i,j\right\rangle ,\sigma }}{%
\delta \left\langle \Psi _{0}\right\vert }+h.c. \notag \\
&&+\sum_{i,\sigma }\frac{%
\partial W}{\partial n_{i,\sigma }}\frac{\delta n_{i,\sigma }}{\delta
\left\langle \Psi _{0}\right\vert }-\lambda \left\vert \Psi
_{0}\right\rangle .  \label{eq:partialW}
\end{eqnarray}%
For an operator $\hat{O}$ with the expectation value \ $O=\left\langle \Psi
_{0}\right\vert \hat{O}\left\vert \Psi _{0}\right\rangle $, $\delta
\left\langle \Psi _{0}\right\vert \hat{O}\left\vert \Psi _{0}\right\rangle
/\delta \left\langle \Psi _{0}\right\vert =\hat{O}\left\vert \Psi
_{0}\right\rangle $. Thus one obtains the following mean field
Hamiltonian,
\begin{eqnarray}
H_{MF}&=&\sum_{\left( i,j\right) ,\sigma }\frac{\partial W}{\partial \chi
_{\left( i,j\right) ,\sigma }}\hat{c}_{i,\sigma }^{\dag }\hat{c}_{j,\sigma
}+h.c. \notag \\
&&+\sum_{\left\langle i,j\right\rangle ,\sigma }\frac{\partial W}{\partial
\Delta _{\left\langle i,j\right\rangle ,\sigma }}\sigma \hat{c}_{i,\sigma }%
\hat{c}_{j,\overline{\sigma}}+h.c. \notag \\
&&+\sum_{i,\sigma }\frac{\partial W}{\partial
n_{i,\sigma }}\hat{n}_{i,\sigma },  \label{eq:variationH}
\end{eqnarray}%
which satisfies the Schr\"{o}dinger equation $H_{MF}\left\vert \Psi
_{0}\right\rangle =\lambda \left\vert \Psi _{0}\right\rangle $. The
coefficients of $H_{MF}$ are given as
\begin{eqnarray}
\frac{\partial W}{\partial \chi _{\left( i,j\right) ,\sigma }}
&=&-\delta _{\left( i,j\right) ,\langle i,j\rangle }J_{\langle
i,j\rangle }\left( \frac{g_{\langle i,j\rangle
}^{s,z}}{4}+\frac{g_{\langle i,j\rangle }^{s,xy}}{2}\frac{\chi
_{\langle i,j\rangle ,\overline{\sigma}}^{\ast }}{\chi _{\langle
i,j\rangle
,\sigma }^{\ast }}\right) \chi _{\langle i,j\rangle ,\sigma }^{\ast } \notag \\
&&-g_{\left(
i,j\right) ,\sigma }^{t}t_{\left( i,j\right) }+\left[
\frac{\partial W}{\partial \chi _{\left( i,j\right) ,\sigma }}\right] _{g};
\notag \\
\frac{\partial W}{\partial \Delta _{\langle i,j\rangle ,\sigma }}
&=&-J_{\langle i,j\rangle }\left( \frac{g_{\langle i,j\rangle }^{s,z}}{4}+%
\frac{g_{\langle i,j\rangle }^{s,xy}}{2}\frac{\Delta _{\langle i,j\rangle ,%
\overline{\sigma}}^{\ast }}{\Delta _{\langle i,j\rangle ,\sigma
}^{\ast }}\right)
\Delta _{\langle i,j\rangle ,\sigma }^{\ast } \notag \\
&&+\left[ \frac{\partial W}{%
\partial \Delta _{\langle i,j\rangle ,\sigma }}\right] _{g}; \notag
 \\
\frac{\partial W}{\partial n_{i,\sigma }} &=&-\left( \mu -V_{i}\right) +%
\frac{1}{2}\sigma \sum_{j}g_{\langle i,j\rangle
}^{s,z}J_{\langle i,j\rangle }m_{j} \notag +\left[ \frac{\partial W}{\partial
n_{i,\sigma }}\right] _{g}  \notag\\
\label{eq:coefficient}
\end{eqnarray}%
with $\partial W/\partial \chi _{\left( i,j\right) ,\sigma }^{\ast }=\left[
\partial W/\partial \chi _{\left( i,j\right) ,\sigma }\right] ^{\ast }$, $%
\partial W/\partial \Delta _{\left( i,j\right) ,\sigma }^{\ast }=\left[
\partial W/\partial \Delta _{\left( i,j\right) ,\sigma }\right] ^{\ast }$, $%
\delta _{\left( i,j\right) ,\langle i,j\rangle }=1$ only when i and j are
nn, otherwise it equals 0, the partial derivative terms $\left[ \frac{%
\partial W}{\partial O}\right] _{g}$ in the above equations refer to the derivative of $W$ with respect to the mean field $O$
via the Gutzwiller g-factors (see Eq[\ref{eq:gfactors}]). This
mean field Hamiltonian $H_{MF}$ in Eq[\ref{eq:variationH}] is then solved
self-consistently. In the numerical calculations, we always
diagonalize $H_{MF}$ for a sample consisting of 257 supercells
along the direction with periodic boundary condition unless stated
explicitly otherwise.

\section{Simplified Model: Site-centered Anti-Phase Domain Walls with d-wave Superconductivity
}

\begin{figure}[t]
\includegraphics
[width=7.0cm,height=16.0cm,angle=270]
{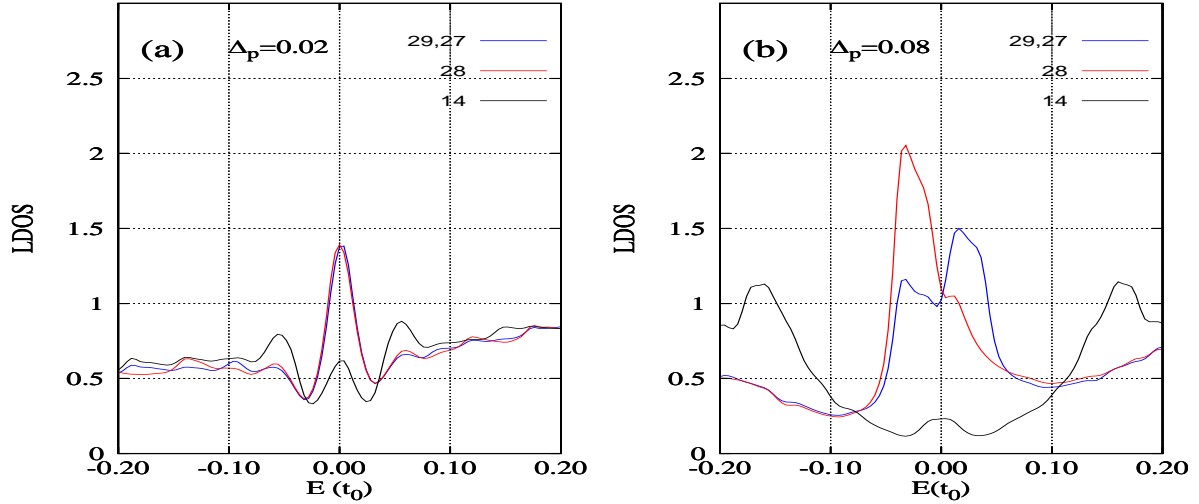}

\caption{(Color online) Local density of states (LDOS) for a simplified model (Hamiltonian
$H_{s}$ defined in Eq[\protect\ref{eq:Htoy}]) for an isolated site-centered anti-phase domain wall in a d-wave SC. Periodic boundary condition along y direction and open boundary condition along x direction are
imposed. The width of the system along x direction is $L_{x}=55$ with the domain wall located at site 28. The average doping
concentration is fixed at $\protect\delta =0.25$, and
$\widetilde{\protect\chi}_{p}=0.20$. Panels (a) and (b) are for
$\widetilde{\Delta}_{p}=0.02$, and $0.08$ respectively. The 14th
site is halfway between the domain wall (site 28) and the
edge, and the
two sites (27, 29) are neighbors of the domain wall. A broadening factor $%
0.004t_{0}$ is used.}
\label{fig:toy}
\end{figure}

\begin{figure}[t]
\includegraphics
[width=18.0cm,height=13.0cm,angle=00]
{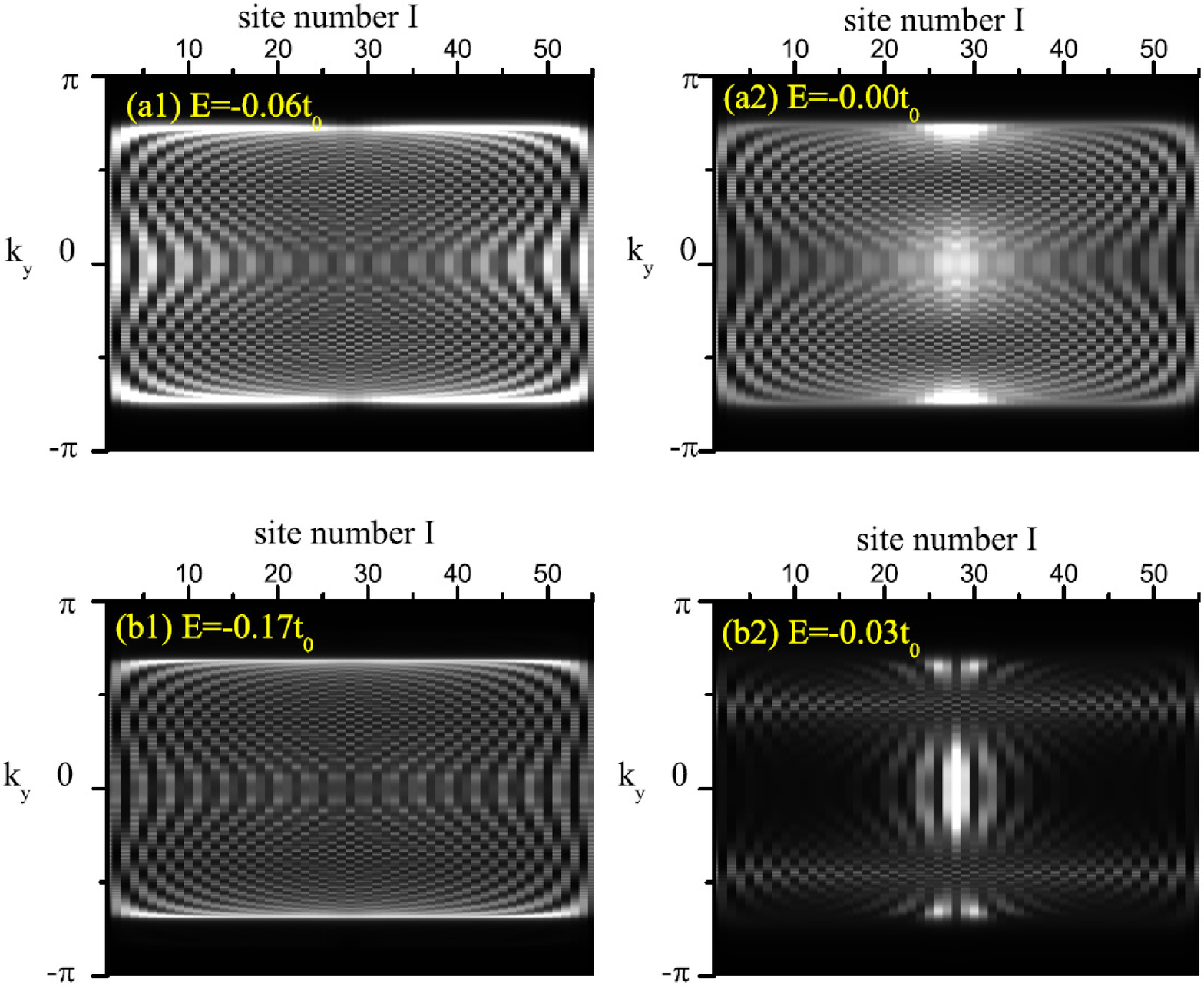}

\caption{(Color online) The spatial (I) and wavevector ($k_{y}$) dependence of the quasiparticle
spectral weight $A_{I,k_{y}}(E)$ for the simplified model ($H_{s}$ in Eq[%
\protect\ref{eq:Htoy}]) with an isolated site-centered anti-phase domain wall in a d-wave SC. The parameters are the same as that used in Fig[\protect\ref{fig:toy}%
]. Panels (a) and (b) are for $\widetilde{\Delta}_{p}=0.02$
($0.08$), respectively. The energies $E$ corresponds to the
Andreev bound states (ABS) in the r.h.s.
panels and the bulk SC gap in the l.h.s. panels as shown in Fig[%
\protect\ref{fig:toy}a,b]. In panel (a2) ABS extends away from the domain wall at site 28 into the bulk of the superconducting state due to $%
\left\vert \Delta \right\vert /E_{F}<<1$, while in panel(b2) where $%
\left\vert \Delta \right\vert $ is much larger the ABS is much more confined
in a small region around the domain wall. For the states close to the SC gap, small $\Delta $ leads to a more homogeneous state,
while moderate $\Delta $ results in a great suppression of the state close
to domain wall.}
\label{fig:toy_ABS}
\end{figure}

We begin the discussion of the results with the simplest case namely
site-centered anti-phase domain walls in a d-wave superconductor (APdSC$^{s}$). To
this end we restrict the Hamiltonian to the two terms without SDW order, and
solve it self-consistently without considering explicitly the doping
dependence of g-factors,
\begin{eqnarray}
H_{s}&=&-\sum_{\left\langle i,j\right\rangle ,\sigma }\left(
g^{t}t_{0}+g^{s}J_{0}\tilde{\chi}_{i,j}^{\ast }\right)
\hat{c}_{i,\sigma }^{\dag
}\hat{c}_{j,\sigma } \notag \\
&&-\sum_{\left\langle i,j\right\rangle }g^{s}J_{0}\tilde{\Delta}%
_{i,j}^{\ast }\hat{c}_{i,\uparrow }^{\dag }\hat{c}_{j,\downarrow
}^{\dag }+h.c. \label{eq:Htoy}
\end{eqnarray}%
Note that $\tilde{\chi}_{i,j}\neq \chi _{i,j}$, and $\widetilde{\Delta}%
_{i,j}\neq \Delta _{i,j}$ but has the same symmetry as $\Delta
_{i,j}$. To keep the model simple, we set
$\widetilde{\chi}_{i,j}=\widetilde{\chi}_{p}$ independent of
$\left\langle i,j\right\rangle $, and $g^{t}=2\delta /\left(
1+\delta \right) $, $g^{s}=4/\left( 1+\delta \right) ^{2}$ where
$\delta $
is the average doping away from half-filling. We consider first an isolated $%
\pi $DW which lies in the center ($i_{x}=28$) of a finite sample
with open boundary condition along x direction and width
$L_{x}=55$. To this end we set $\left\vert
\widetilde{\Delta}_{i,j}\right\vert =\widetilde{\Delta}_{p}$
except for the bonds
along the domain wall which are set to zero, i.e $\widetilde{\Delta}%
_{i,j}|_{i_{x}=j_{x}=28}=0$. The $\pi $-phase shift requires that for the
two bonds $\left\langle i,j\right\rangle $ and $\left\langle i^{\prime
},j^{\prime }\right\rangle $ which are located symmetrically on the two sides of
the domain wall, $\Delta _{i,j}|_{i_{x},j_{x}\leq 28}=-\Delta _{i^{\prime
},j^{\prime }}|_{i_{x}^{\prime },j_{x}^{\prime }\geq 28}$. The change of
sign at the domain wall causes an Andreev bound state (ABS) to appear at the
chemical potential which we take as the energy zero. This shows up clearly
when we calculate the local density of states (LDOS) as illustrated in Fig[%
\ref{fig:toy}a,b]. For the case of weak coupling in
Fig[\ref{fig:toy}a] a clear peak appears in the LDOS at zero
energy for sites at or near the domain wall ($i_{x}=27,28,29$)
while far away sites show peaks at the bulk gap edges and reduced
values at zero energy. This behavior shows up also very clearly in
the spatial dependence of the quasiparticle spectral weight. This
is illustrated in Fig[\ref{fig:toy_ABS}a,b] for the case of a weak and a
moderate gap value of the pairing amplitude
$\widetilde{\Delta}_{p}=0.02$($0.08$). The spectral weight is
concentrated close to the $\pi $DW at quasiparticle energies
$E_{k}\simeq 0$, but away from the $\pi $DW for values of $E_{k}$
near the bulk gap energy
$E_{k}=2g^{s}J_{0}\widetilde{\Delta}_{p}$. The total
energy differences between the states with and without $\pi $DW for the two $\widetilde{\Delta}_{p}$  are $0.0066t_{0}$ and $0.0365t_{0}$, respectively. The energy cost of the domain
wall
is substantial, consistent with the creation of a LDOS peak in the center of
energy gap. Note that for the case of a moderate gap value of $\widetilde{%
\Delta}_{p}$, the peak of LDOS near $E_{k}\simeq 0$ shows structures consistent with
the development of a one-dimensional band of Andreev bound states which
propagate along the domain wall. This can be also seen in the quasiparticle
dispersion which is a function only of $k_{y}$. \newline

\begin{figure}[tbp]
\begin{minipage}[t]{0.5\linewidth}
\centering
\includegraphics[width=9.0cm,height=7.0cm, angle=0]
{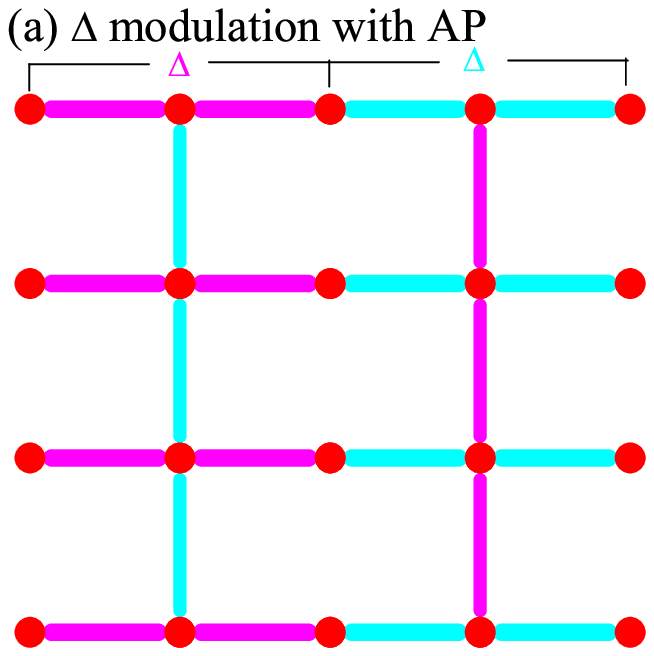}
\end{minipage}%
\begin{minipage}[t]{0.5\linewidth}
\centering
\includegraphics[width=7.0cm,height=6.5cm, angle=0]
{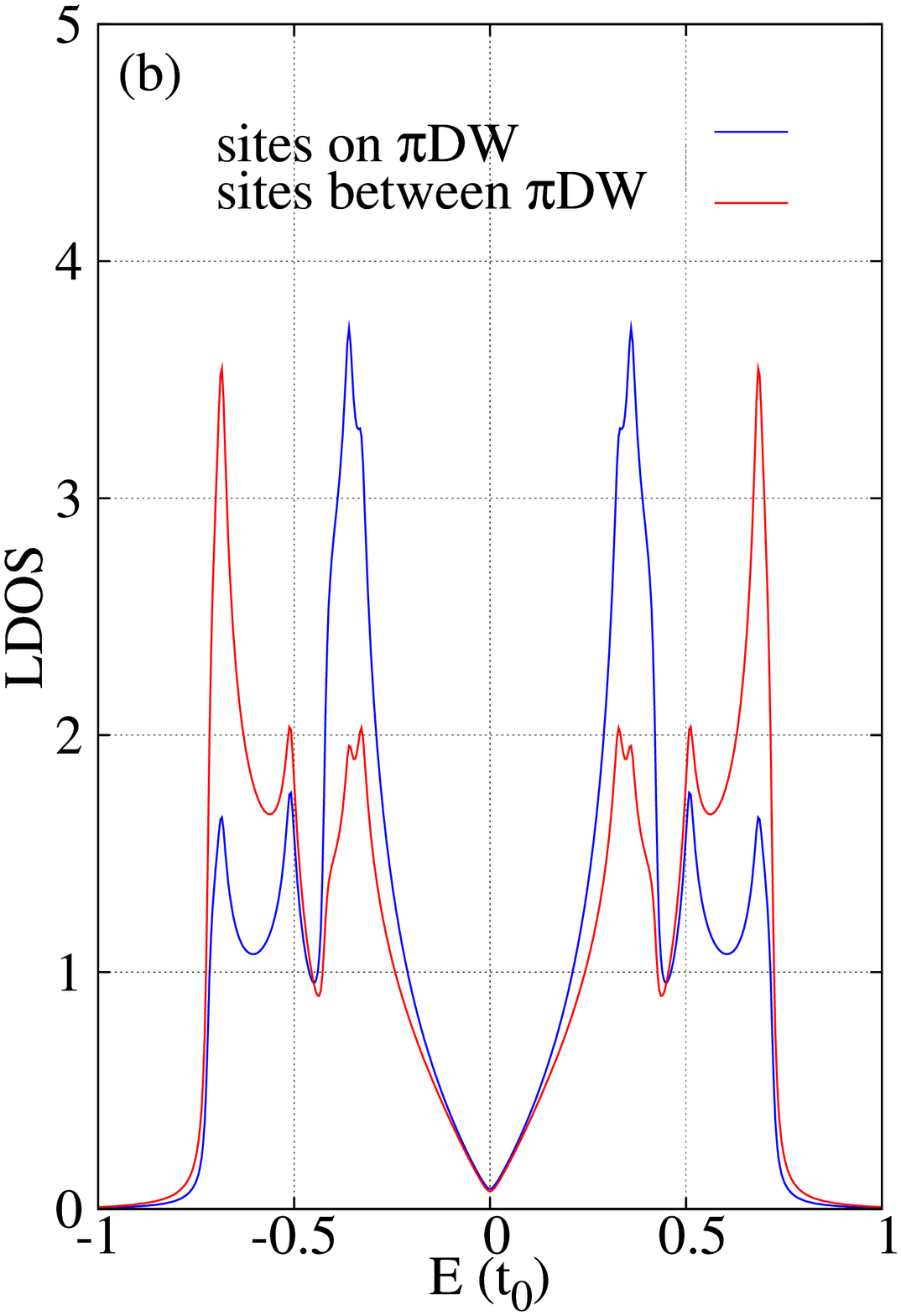}
\end{minipage}
\caption{(Color online) (a) Schematic illustration of the
modulation of the pairing parameter $ \Delta$ for the simplified
model (Hamiltonian $H_{s}$ defined in Eq[\ref{eq:Htoy}]) for dSC
state with periodic site-centered anti-phase domain walls with the
shortest periodicity of $L_{x}=4$. The anti-phase pattern of
$\Delta$ is illustrated by the color scheme. (b) Local
density of states (LDOS) with parameter values as doping concentration $\protect\delta %
=0$ and $\widetilde{\protect\chi}_{p}=\widetilde{\Delta}_{p}=0.2$. The domain walls are close to each other, they form bands with weak dispersion along $%
k_{x}$ but strong dispersion along $k_{y}$ parallel to the domain wall. At half filling, these bands display an
anisotropic nodal structure as demonstated in
Eq[\protect\ref{eq:node_PDRVB4L}] and by the low energy LDOS
behavior.} \label{fig:PDRVB_L4}
\end{figure}

Turning our attention to a periodic array of parallel $\pi $DW, we focus on
the case of period $L_{x}=4$, relevant to the cuprates, illustrated in Fig[%
\ref{fig:PDRVB_L4}]. In this case the Andreev bound states on neighboring
domain walls will overlap strongly leading to a more complex dispersion
relation for the associated quasiparticle states. Note the d-wave form of
the bulk superconductivity leads to gapless excitations in the nodal
directions which in turn leads to stronger overlap for near nodal
quasiparticles. To illustrate this more complex behavior we focus on a
particular model which can be solved analytically. To this end we set $%
\delta =0$ (i.e. half-filling), $g^{t}=0$ and set $\widetilde{\chi}_{p}=\widetilde{%
\Delta}_{p}$ and $g^{s}J\widetilde{\chi}_{p}=1$. In this case the
quasiparticle dispersion is obtained by diagonalizing the
Hamiltonian
\begin{eqnarray}
H_{k} &=&-\mathbf{X}_{k}^{\dag }\left(
\begin{array}{cc}
A_{k} & B_{k} \\
B_{k}^{\ast } & -A_{-k}^{\ast }\label{Eq:Ham_PDRVB4L}%
\end{array}%
\right) \mathbf{X}_{k}; \\
A_{k} &=&\left(
\begin{array}{cccc}
2\cos k_{y} & e^{ik_{x}} & 0 & e^{-ik_{x}} \\
e^{-ik_{x}} & 2\cos k_{y} & e^{ik_{x}} & 0 \\
0 & e^{-ik_{x}} & 2\cos k_{y} & e^{ik_{x}} \\
e^{ik_{x}} & 0 & e^{-ik_{x}} & 2\cos k_{y}%
\end{array}%
\right) ; \notag \\B_{k}&=&\left(
\begin{array}{cccc}
0 & e^{ik_{x}} & 0 & -e^{-ik_{x}} \\
e^{-ik_{x}} & -2\cos k_{y} & e^{ik_{x}} & 0 \\
0 & e^{-ik_{x}} & 0 & -e^{ik_{x}} \\
-e^{ik_{x}} & 0 & -e^{-ik_{x}} & 2\cos k_{y}%
\end{array}%
\right),\notag
\end{eqnarray}%
where $\mathbf{X}_{k}^{\dag }=\left( \hat{c}_{I,k,\uparrow }^{\dag
},\hat{c}_{I,-k,\downarrow }\right) $ with $I=1,2,3,4$ denoting
the sites inside a supercell. The quasiparticle dispersion takes a
simple form,
\begin{eqnarray}
E_{k}=\pm \sqrt{6\cos ^{2}k_{y}+4\pm 2\sqrt{\left( 2+\cos ^{2}k_{y}\right)
^{2}-4\sin ^{2}2k_{x}}}.\notag \\  \label{eq:Dis_PDRVB4L}
\end{eqnarray}%
For a wavevector $\left( k_{x},k_{y}\right) $ close to $\left( \pi /2,\pi
/2\right) $, the two quasiparticle bands close to the Fermi level have an
anisotropic nodal structure with
\begin{equation}
E_{k}=\pm 2\sqrt{K_{y}^{2}+2K_{x}^{2}},  \label{eq:node_PDRVB4L}
\end{equation}%
where $\left( K_{x},K_{y}\right) =\left( \pi /2-k_{x},\pi /2-k_{y}\right) $.
This nodal structure completely suppresses the the density of states (DOS)
at zero energy as shown in Fig[\ref{fig:PDRVB_L4}], and pushes the peaks in
the DOS of the Andreev bound states away from the chemical potential.

\section{Coexisting Anti-Phase Superconductivity and Spin and Charge Density
Waves}

Anti-phase domain walls in a superconductor usually cost a
substantial energy. The key question raised by the recent
experimental results of Li \textit{et al} \cite{Li-PRL-07}. on the
static stripe phase is whether SDW and CDW coexisting with $\pi
$DW lead to a state with a net energy gain. The VMC calculations
of Himeda \textit{et al.} \cite{Himeda-PRL-02} found a small
energy gain for a longer superlattice with a larger separation
between $\pi $DW within a restricted parameter range. Recent
calculations for a 8-superlattice without SDW order by Raczkowski
\textit{et al.} \cite{Raczkowski-PRB-07} did not yield an energy
gain but the energy cost to introduce $\pi $DW was quite small.
These results motivated us to examine a wider parameter range
within a RMFT approach and look for a possible net energy gain in
an 8-superlattice (with site-centered anti-phase domain walls) at a hole concentration $\delta
=1/8$ when coexisting SDW order and $\pi $DW
are included. A longer 10-superlattice (with bond-centered anti-phase domain walls) state gives similar results.
 In view of the orthorhombic nature of the individual CuO$_{%
\text{2}}$-planes in the LTT phase, we allowed for anisotropy in
the hopping $t_{x(y)}$ and exchange coupling $J_{x(y)}$. Below we
keep the nn hopping in
the y-direction fixed, $t_{y}=t_{0}$, and scale $%
J_{x}/J_{y}=t_{x}^{2}/t_{y}^{2}$. In addition the presence of a
crystallographic superlattice in the LTT phase motivated us to
examine also the effect of the lattice inhomogeneity by including
a site dependent potential modulation, $V_{i}$.

\subsection{Site-centered anti-phase dSC}

\begin{figure}[tbp]
\begin{minipage}[t]{0.5\linewidth}
\centering
\includegraphics[width=3.5in, angle=0]
{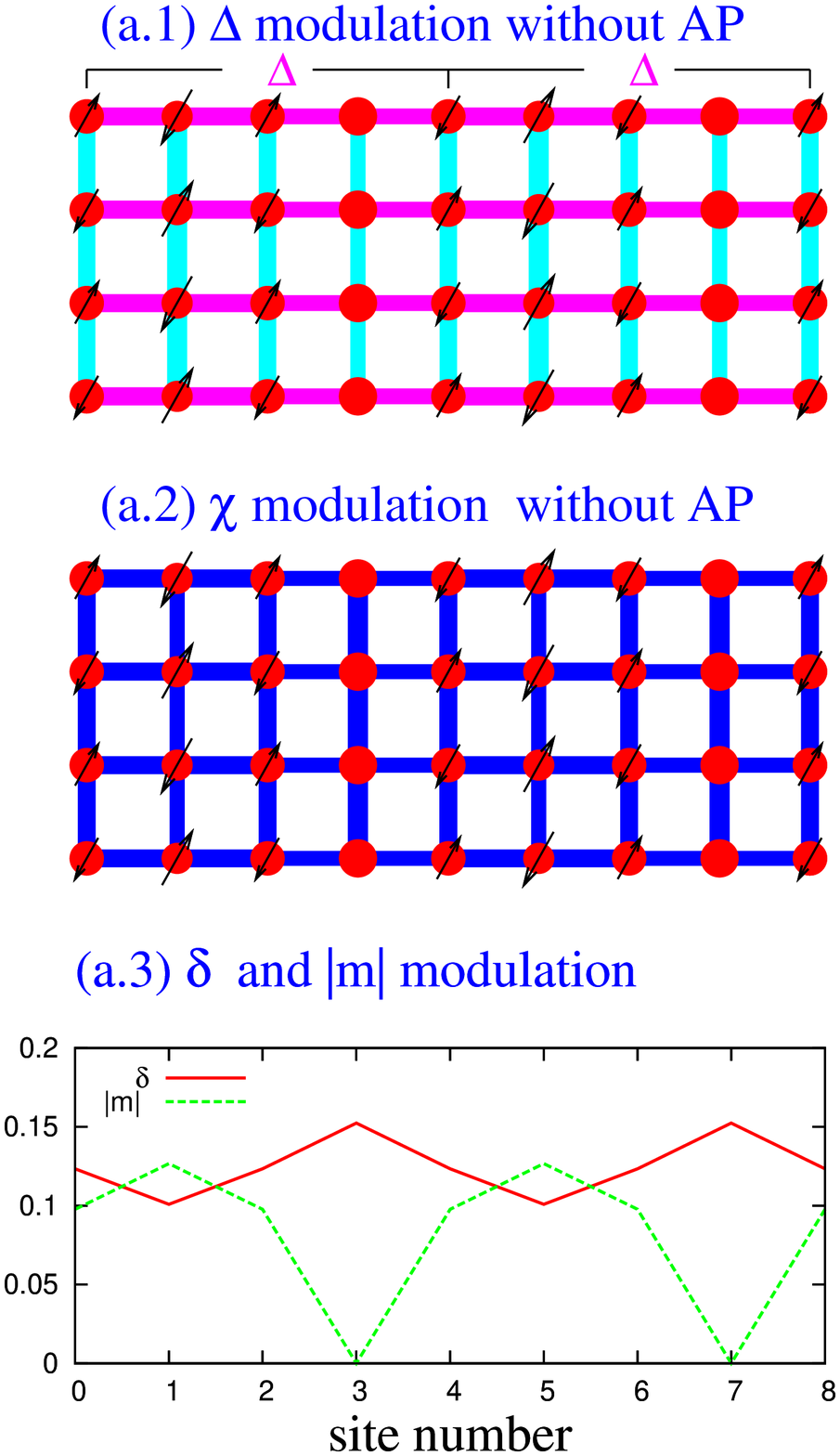}
\end{minipage}%
\begin{minipage}[t]{0.5\linewidth}
\centering
\includegraphics[width=3.5in, angle=0]
{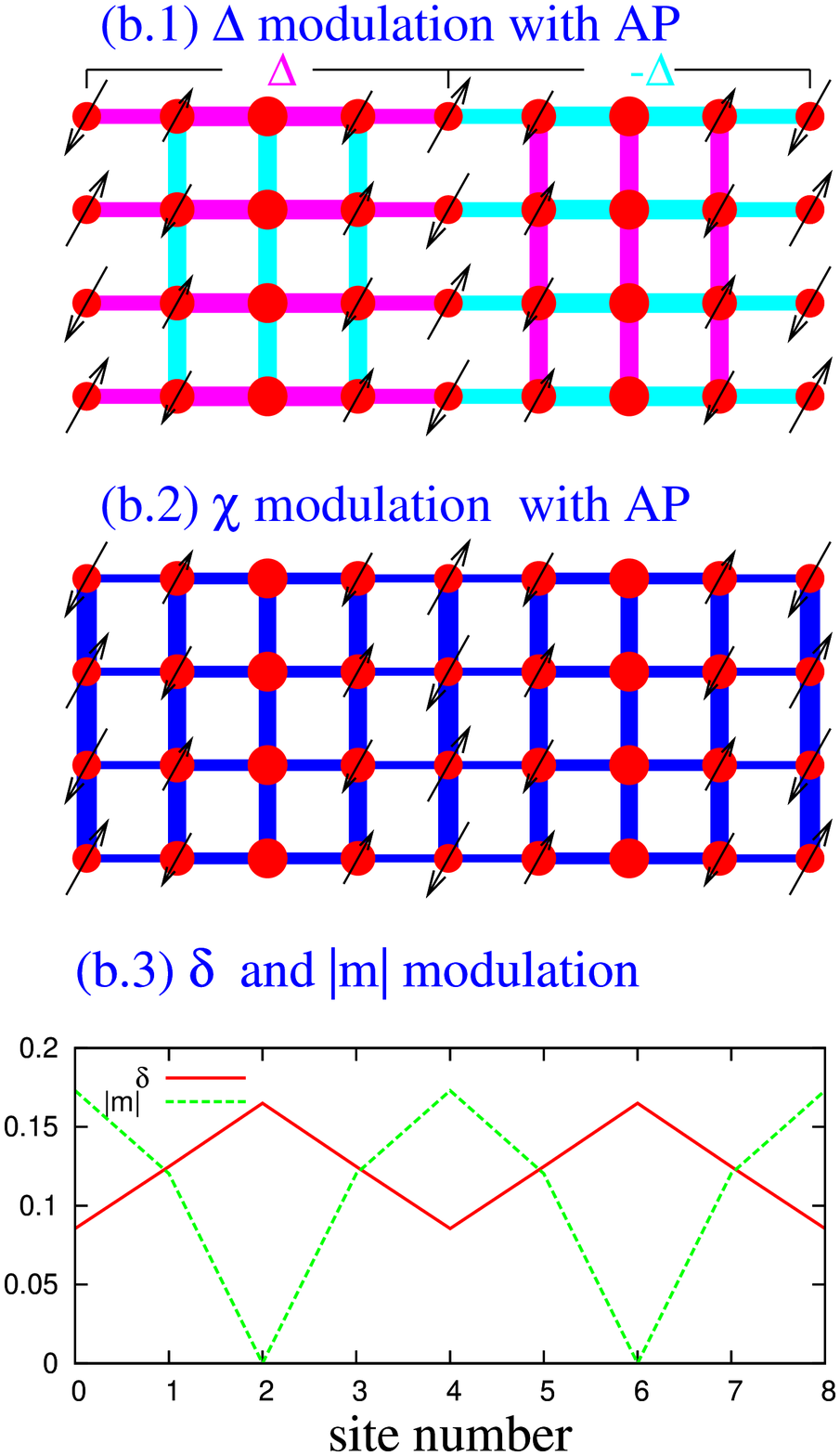}
\end{minipage}
\caption{(Color online) Schematic illustration of the modulations of the parameters pairing
amplitude $\Delta $, hopping amplitude $\protect\chi $, hole concentration $%
\protect\delta $ and antiferromagnetic moment $m$, for two states:
SDW+CDW+dSC$^{s}$ [panels (a)] and SDW+CDW+APdSC$^{s}$ [panels (b)] (without and with site-centered anti-phase domain walls). The
average doping
is $1/8$ and the periodicity $L_{x}=8$. In panels (a/b 1-2) The amplitudes $%
\Delta $ and $\protect\chi $ are denoted by the width of the bond, the
spatial modulation of the staggered antiferromagnetic moment $m_{i}$ is
denoted by the arrows, the hole concentration modulation is represented by
the size of the dots. The anti-phasing of $\Delta $ in panel(b.1) is
illustrated by the different color pattern at either side of the domain wall with cyan (magenta) for positive (negative) value. D-wave pairing
symmetry is still preserved between two neighboring domain walls. The
anti-phase domain walls coincidence with the sites which have the largest sublattice
antiferromagentic moment and smallest hole concentration. However, for the
case without SDW, the domain walls locate at the sites with the largest
hole concentration. \protect\cite{Raczkowski-PRB-07} Panels (a/b 3) show the
spatial hole density (red solid) and the AF moment (green dash) modulations. The site-centered anti-phase domain walls lead to an anisotropy of $\protect\chi $, and an enhancement
of the hole density and antiferromagnetic moment modulation. }
\label{fig:DWconfig}
\end{figure}

The RMFT approximation yields a series of coupled nonlinear
equations. An iteration method is used to obtain optimal values of
the four order parameters: the pairing and hopping amplitudes,
sublattice magnetization and hole density. When the solution
iterates to a stable set of values we can conclude that a local
energy minimum exists, but on occasion no stable solution can be
found, which indicates that no local minimum exists with this
symmetry. In general we find stable solutions for the case of
coexisting CDW and SDW with or without $\pi $DW. Typical patterns
for an 8-superlattice are illustrated in Fig[\ref{fig:DWconfig}]
with or without site-centered $\pi $DW in systems where the
modulation of the pairing amplitude is site centered. The
antiferromagnetic domain wall (AFDW) coincides with the maximum
hole density while the $\pi $DW appears at the minimum hole
density. (In the case without SDW, the $\pi $DW appears at the
maximum hole density \cite{Raczkowski-PRB-07}.)

\begin{figure}[b]
\includegraphics
[width=12.0cm,height=12.0cm,angle=0]
{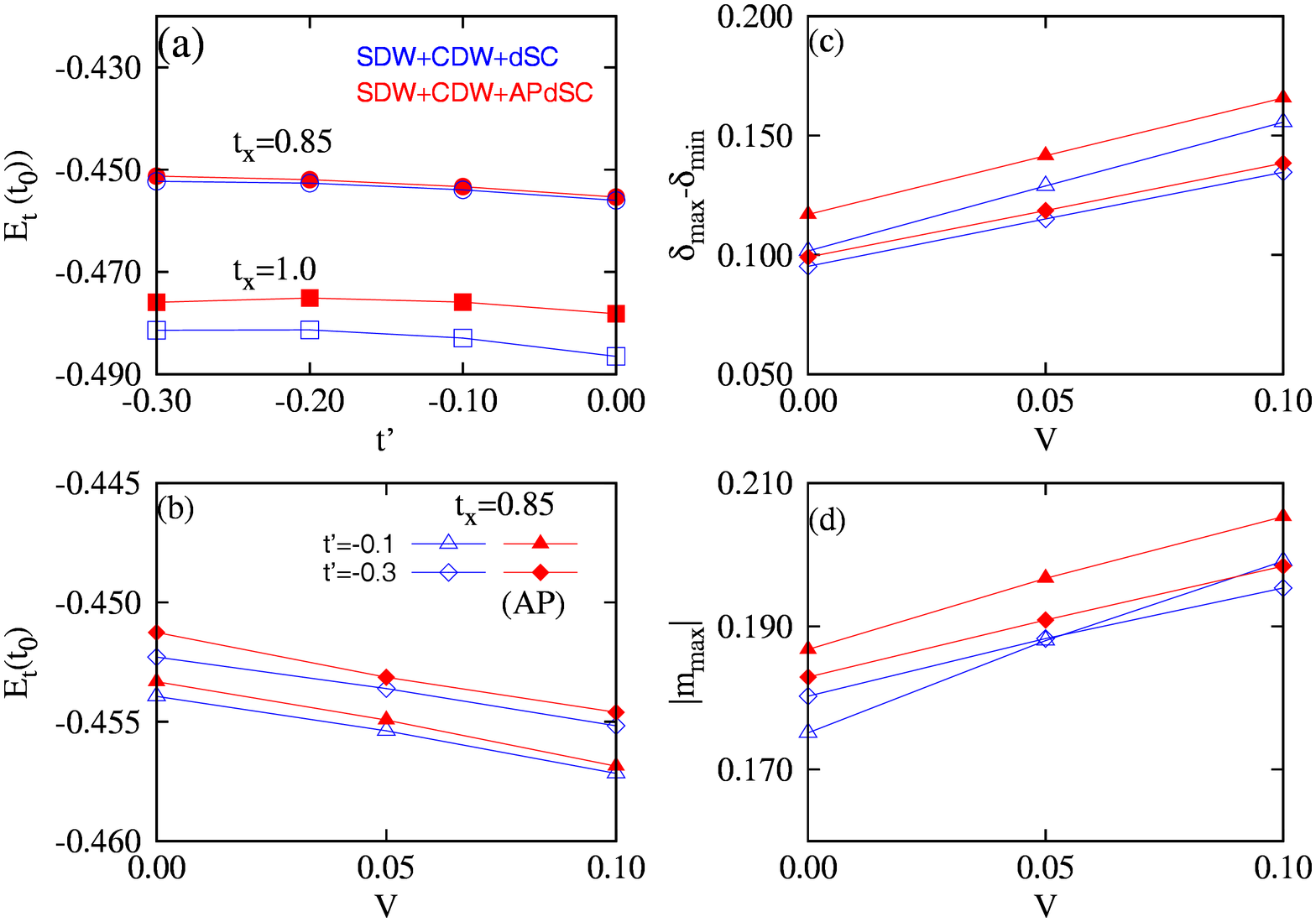}

\caption{(Color online) (a) The energy (shown in Eq[\ref{eq:energy}]) dependence of the two
states SDW+CDW+dSC$^{s}$ and SDW+CDW+APdSC$^{s}$ (without and with site-centered anti-phase domain walls) on the nnn hopping integral
$t^{\prime }$ for isotropic and
anisotropic nn hopping ratio $t_{x}/t_{y}$. The energy unit is $t_{0}=300meV$. The nnn hopping integral $%
t^{\prime }$ does not but anisotropic $t_{x\left( y\right) }$ and $%
J_{x\left( y\right) }$ do help to push the energy of SDW+CDW+APdSC$^{s}$
state (the solid and red symbol) closer to SDW+CDW+dSC$^{s}$ state (the
open and blue
symbol). Square (circle) symbols are for the values $t_{x}/t_{y}=1.00 (0.85)$%
. Panels (b, c, d): the energy, charge and magnetization moment
modulations of these two states with additional external
potentials which are imposed to enhance the charge and magnetic
modulations by shifting the local potential by $+V$ $(V>0) $ up
for the sites with the largest antiferromagnetic moment and  $-V$
down for the sites with zero antiferromagnetic moment. A
substantial anisotropy $t_{x}/t_{y}=0.85$ is used. The diamond
(triangle) symbols are for $t^{\prime}=-0.3 (-0.1)$. The larger
antiferromagnetic moment and hole concentration modulations drive
the energy difference smaller between the two states. }
\label{fig:energy_comparison}
\end{figure}

In table \ref{table:energy} the results for the ground state
energy and local values of the order parameters are presented. The
upper lines are for the case of nn hopping only ($t^{\prime }=0$),
with and without, anisotropic component in the nn hopping
$t_{x(y)}$. In this case $t_{x}=t_{y}$ the results show that the
uniform AFM+dSC state is lowest. When AFDW and the associated
modulation of the hole density are included the resulting state
(denoted by SDW+CDW+dSC$^{s}$) has an energy that is slightly higher. Introducing $%
\pi $DW to create antiphase superconducting order (SDW+CDW+APdSC$^{s}$) raises the
energy further. Anisotropy in the nn hopping narrows the energy differences
but does not change the relative ordering of the states with and without $%
\pi $DW. When a weak nnn hopping is added, the SDW+CDW+dSC$^{s}$ state gains in
energy and when anisotropy is also added this state has the lowest energy.
In this case when we consider anisotropic nn hopping, the energy cost of
introducing $\pi $DW in the superconducting is further reduced to small but
still positive values. A further increase in the nnn hopping term (shown in
Fig[\ref{fig:energy_comparison}a]) however does not lead to an energy gain
for $\pi $DW. The energy cost of $\pi $DW remains very small but positive.

The presence of substantial local modulations in the hole density
in these states led us to investigate the effect of introducing a
site dependent potential shift. Such a shift can result from the
crystallographic superlattice modulation that appears at the
crystallographic transition into the LTT state. The results in
Fig[\ref{fig:energy_comparison}b] show that this potential shift
reduces the energy cost of the site-centered anti-phase domain wall and enhances the charge
and spin modulation but still does not lead to a net energy gain
for the SDW+CDW+APdSC$^{s}$ state even in the most favorable case of anisotropic
nn hopping and substantial nnn hopping. Within the RMFT the $\pi
$DW always demands an energy cost even though it may be only a
very small amount. Bond-centered $\pi $DW with anisotropic nn
hopping and longer periodicity $L_{x}=10$ shows that the energy
difference between these two states, with and without $\pi$DW, can
be also very close.

\begin{widetext}
\begin{center}
\begin{table}[tbp]
\tiny
\begin{tabular}{|l|l|l|l|l|l|l|l|l|l|l|l|l|}
 \hline\hline
$t_{\text{x}},t^{\prime} $ & state & $E_{t}$ & $E_{\text{kin}}$ & $E_{\text{%
J }}$ & $\delta_{\max }$ & $\delta _{\min }$ & $m_{max}$ &
$\overline\Delta_{max}$ & $\overline\Delta_{min}$ &
$\overline\chi_{max}$ & $\overline\chi_{min}$ \\ \hline\hline &
AFM+dSC & -0.4878 & -0.3287 & -0.1593 & 0.12500 & 0.12500 &
0.08524 &
0.1142 & 0.1142 & 0.1903 & 0.1903 \\
\raisebox{0ex}{$t_{x}=1.00$} & dSC & -0.4863 & -0.3428 & -0.1435 & 0.12500 &
0.12500 & 0 & 0.1152 & 0.1152 & 0.1928 & 0.1928 \\
\raisebox{0ex} { $t^{\prime}=0.0$} & SDW+CDW+dSC$^{s}$ & -0.4865 & -0.3373 &
-0.1492 & 0.1372 & 0.1134 & 0.08412 & 0.1214 & 0.09917 & 0.2215 & 0.1821 \\
& SDW+CDW+APdSC$^{s}$ & -0.4782 & -0.3292 & -0.1490 & 0.1498 & 0.09604 &
0.1418 & 0.1114 & 0 & 0.2639 & 0.1111 \\ \hline\hline
& AFM+dSC & -0.4536 & -0.3117 & -0.1419 & 0.12500 & 0.12500 & 0.07432 &
0.08399 & 0.07724 & 0.2652 & 0.1098 \\
\raisebox{0ex}{$t_{x}=0.85$} & dSC & -0.4526 & -0.3225 & -0.1301 & 0.12500 &
0.12500 & 0 & 0.08409 & 0.07593 & 0.2675 & 0.1122 \\
\raisebox{0ex}{$t^{\prime}=0.0$} & SDW+CDW+dSC$^{s}$ & -0.4560 & -0.3050 &
-0.1510 & 0.1737 & 0.07130 & 0.1720 & 0.06538 & 0.05154 & 0.2756 & 0.07822 \\
& SDW+CDW+APdSC$^{s}$ & -0.4554 & -0.3036 & -0.1518 & 0.1815 & 0.05752 &
0.1871 & 0.04479 & 0 & 0.2866 & 0.06392 \\
& SDW+CDW+dSC$^{b}$ & -0.4567 & -0.3002 & -0.1564 & 0.1911 & 0.06029 & 0.1831
& 0.06692 & 0.05151 & 0.2769 & 0.05869 \\
& SDW+CDW+APdSC$^{b}$ & -0.4563 & -0.2978 & -0.1586 & 0.1988 & 0.04986 &
0.1941 & 0.06055 & 0 & 0.2841 & 0.03663 \\ \hline\hline
& AFM+dSC & -0.4841 & -0.3219 & -0.1622 & 0.12500 & 0.12500 & 0.09356 &
0.1149 & 0.1149 & 0.1893 & 0.1893 \\
\raisebox{0ex}{$t_{x}=1.00$} & dSC & -0.4817 & -0.3372 & -0.1445 & 0.12500 &
0.12500 & 0 & 0.1179 & 0.1179 & 0.1920 & 0.1920 \\
\raisebox{0ex}{$t_{x}=-0.1$} & SDW+CDW+dSC$^{s}$ & -0.4829 & -0.3268 &
-0.1561 & 0.1525 & 0.1008 & 0.1268 & 0.1304 & 0.09917 & 0.2215 & 0.1654 \\
& SDW+CDW+APdSC$^{s}$ & -0.4759 & -0.3202 & -0.1557 & 0.1650 & 0.08549 &
0.1731 & 0.09773 & 0 & 0.2642 & 0.1034 \\ \hline\hline
& AFM+dSC & -0.4507 & -0.3056 & -0.1451 & 0.12500 & 0.12500 & 0.08474 &
0.07982 & 0.07170 & 0.2688 & 0.1031 \\
\raisebox{0ex}{$t_{x}=0.85$} & dSC & -0.4490 & -0.3182 & -0.1308 & 0.12500 &
0.12500 & 0 & 0.08140 & 0.07155 & 0.2721 & 0.1054 \\
\raisebox{0ex}{$t^{\prime}=-0.1$} & SDW+CDW+dSC$^{s}$ & -0.4539 & -0.3024 &
-0.1515 & 0.1750 & 0.07336 & 0.1751 & 0.06401 & 0.04880 & 0.2742 & 0.08047 \\
& SDW+CDW+APdSC$^{s}$ & -0.4533 & -0.3021 & -0.1512 & 0.1801 & 0.06318 &
0.1867 & 0.04286 & 0 & 0.1840 & 0.06953 \\
& SDW+CDW+dSC$^{b}$ & -0.4538 & -0.2991 & -0.1547 & 0.1833 & 0.07206 & 0.1740
& 0.07046 & 0.05567 & 0.2728 & 0.07248 \\
& SDW+CDW+APdSC$^{b}$ & -0.4533 & -0.2972 & -0.1560 & 0.1890 & 0.06202 &
0.1860 & 0.05946 & 0 & 0.2810 & 0.04685 \\ \hline\hline
& AFM+dSC & -0.4813 & -0.3151 & -0.1662 & 0.12500 & 0.12500 & 0.1188 & 0.1086
& 0.1086 & 0.1866 & 0.1866 \\
\raisebox{0ex}{$t_{x}=1.00$} & dSC & -0.4750 & -0.3303 & -0.1446 & 0.12500 &
0.12500 & 0 & 0.1216 & 0.1216 & 0.1899 & 0.1899 \\
\raisebox{0ex}{$t^{\prime}=-0.3$} & SDW+CDW+dSC$^{s}$ & -0.4814 & -0.3213 &
-0.1602 & 0.1709 & 0.09043 & 0.1746 & 0.1263 & 0.07922 & 0.2351 & 0.1280 \\
& SDW+CDW+APdSC$^{s}$ & -0.4760 & -0.3236 & -0.1523 & 0.1700 & 0.09028 &
0.2002 & 0.07064 & 0 & 0.2431 & 0.1266 \\ \hline\hline
& AFM+dSC & -0.4491 & -0.2986 & -0.1506 & 0.12500 & 0.12500 & 0.1127 &
0.06892 & 0.05955 & 0.2683 & 0.09673 \\
\raisebox{0ex}{$t_{x}=0.85$} & dSC & -0.4436 & -0.3122 & -0.1314 & 0.12500 &
0.12500 & 0 & 0.08064 & 0.06819 & 0.2762 & 0.09695 \\
\raisebox{0ex}{$t^{\prime}=-0.3$} & SDW+CDW+dSC$^{s}$ & -0.4523 & -0.3008 &
-0.1515 & 0.1774 & 0.08221 & 0.1883 & 0.06455 & 0.04278 & 0.2681 & 0.08799 \\
& SDW+CDW+APdSC$^{s}$ & -0.4518 & -0.3017 & -0.1501 & 0.1787 & 0.08034 &
0.1822 & 0.03503 & 0 & 0.2768 & 0.07811 \\
& SDW+CDW+dSC$^{b}$ & -0.4513 & -0.2985 & -0.1528 & 0.1789 & 0.09117 & 0.1638
& 0.06976 & 0.04589 & 0.2680 & 0.08756 \\
& SDW+CDW+APdSC$^{b}$ & -0.4506 & -0.2984 & -0.1523 & 0.1806 & 0.08737 &
0.1700 & 0.04898 & 0 & 0.2762 & 0.06648 \\ \hline\hline
\end{tabular}%
\caption{ Key results for various states obtained by selfconsistently solving the Hamiltonian in Eq[\ref{eq:variationH}] with an
average hole density of 1/8. Listed are the mean field energy $%
E_{t}$, kinetic energy $E_{kin}$, spin-spin superexchange energy $E_{J}$ and
the modulation of hole concentration ($\protect\delta _{max}$ and $\protect%
\delta _{min}$), the largest antiferromagnetic moment ($|m|_{max}$), pairing
amplitude $\overline{\Delta}_{max}$ and $\overline{\Delta}_{min}$, and $\overline{\protect%
\chi}_{max}$ and $\overline{\protect\chi}_{min}$ ($\overline{\Delta}=\sum_{\protect%
\sigma}\overline\Delta _{\protect\sigma}$, $\overline{\protect\chi}=\sum_{\protect%
\sigma}\protect\chi _{\protect\sigma }$) for various states
including homogeneous AFM+dSC and dSC states,
SDW+CDW+dSC$^{s}$ ($L_{x}=8$),  SDW+CDW+APdSC$^{s}$ ($L_{x}=8$) and
 SDW+CDW+dSC$^{b}$ ($L_{x}=5$), SDW+CDW+APdSC$^{b}$ ($L_{x}=10$)
states [APdSC (dSC) stands for d-wave SC state with (without) anti-phase domain walls site-centered ($^{s}$) or bond-centered ($^{b}$)].  The energy unit is $t_{0}=300meV$, $V\equiv 0$, $J_{x}/J_{y}=t_{x}^{2}/t_{y}^{2}$, $%
t_{y}=1$, $J_{y}=0.3$. Anisotropic nn hopping tends to
energetically favor the homogeneous AFM+dSC state compared to the
homogeneous dSC state. However, it also causes some inhomogeneous
state to be energetically more favored, here for instance, the
SDW+CDW+dSC$^{s}$ state. Note that to introduce anti-phase domain wall in the
pairing order parameter in the renormalized mean field theory for the $t-J$ model always
cost energy, although it can be very small.} \label{table:energy}
\end{table}
\end{center}
\end{widetext}

\subsection{Bond-centered anti-phase dSC}

\begin{figure}[tbp]
\begin{minipage}[t]{0.5\linewidth}
\centering
\includegraphics[width=10.0cm,height=12.0cm, angle=0]
{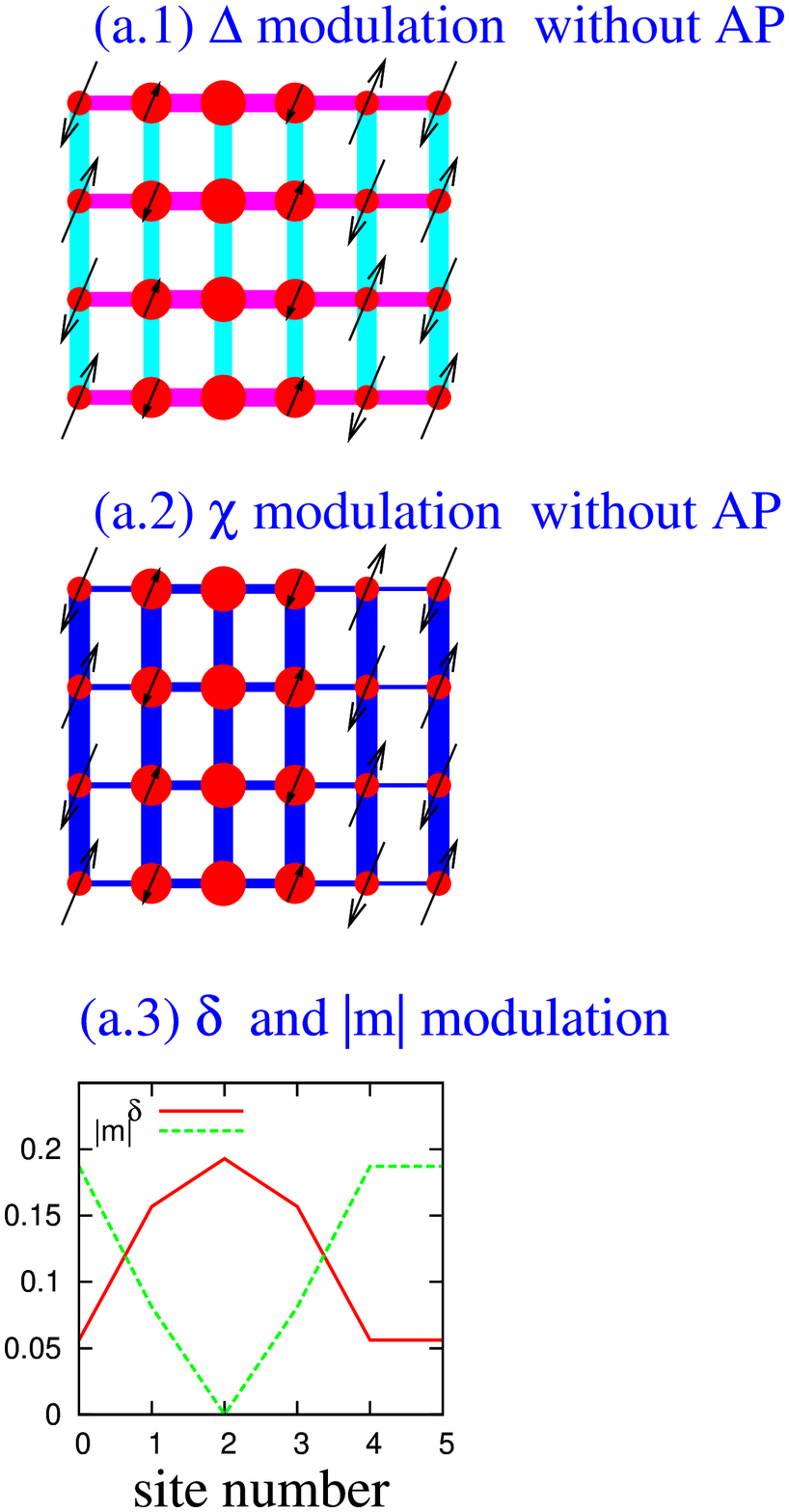}
\end{minipage}%
\begin{minipage}[t]{0.5\linewidth}
\centering
\includegraphics[width=10.0cm,height=12.0cm, angle=0]
{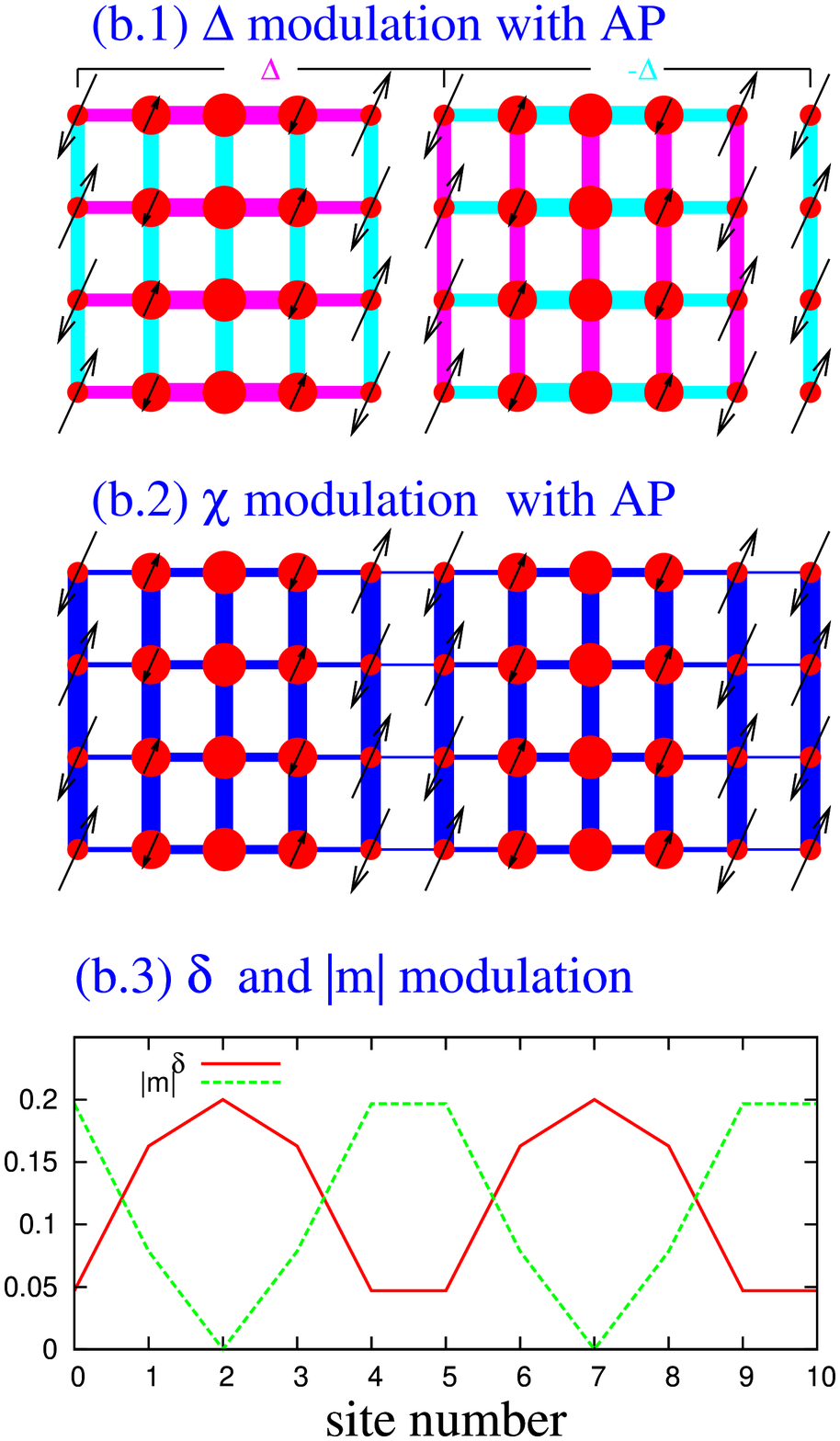}
\end{minipage}
\caption{(Color online) Schematic illustration of the modulations of the parameters pairing
amplitude $\Delta $, hopping amplitude $\protect\chi $, hole concentration $%
\protect\delta $ and antiferromagnetic moment $m$, for two states:
SDW+CDW+dSC$^{b}$ [panels (a)] and SDW+CDW+APdSC$^{b}$ [panels (b)] (without or
with bond-centered anti-phase domain walls). The average doping
is $1/8$ and the periodicity $L_{x}=10$. As shown in panel (b1)
the anti-phase modulation of the pairing $\Delta$ is bond-centered
with the domain wall located at the bonds connecting two nn
sites with maximum stagger antiferromagnetic moment $\pm |m|$ along x
direction. The energy difference between the two states with and
without bond-centered domain wall is even smaller than the
case for site-centered domain wall with anisotropic nn hopping
$t_{x}/t_{y}=0.85$. The modulation magnitude of the hole density and
antiferromagnetic moment in these two states are close to each
other.} \label{fig:BDWconfig}
\end{figure}

Alternative bond-centered anti-phase modulations of the pairing amplitude
were considered by several groups.\cite{Himeda-PRL-02, Raczkowski-PRB-07,
Capello-PRB-08} In the case of the 8-superlattice we did not find any
stable bond centered solution with nonzero SDW in the doping
regime around 1/8 when requiring there is antiferromagnetic domain wall ($m_{I}=0$). But for longer periodicity $L_{x}$=10 we found
a stable solution. In Fig[\ref{fig:BDWconfig}] a typical pattern for
this long 10-superlattice with and without the bond-centered $\pi$DW is
illustrated. The energy cost of the APdSC$^{b}$ is also positive for the
bond-centered case but is even smaller compared with the
site-centered case (see table \ref{table:energy}) at some cases.

\section{Spectral Properties of the Modulated Phases}

Next we examine the density of states in the modulated phases
which gives us insight into the interplay between the SDW and SC with
either dSC or APdSC order in the stripe phases. We restrict our
considerations to the case of site-centered pairing modulation
relevant for 8-superlattice. It is instructive to calculate
several density of states, starting with the local density of
states (LDOS)
\begin{equation}
A_{I}\left( \omega \right) =-\frac{1}{\pi }\sum_{\sigma }ImG_{I,\sigma
}\left( \omega \right),  \label{LDOS}
\end{equation}%
where $G_{I,\sigma }\left( \omega \right) $ is the Fourier transform of the
time dependent onsite Green's function $G_{I,\sigma }\left( t\right)
=-i\left\langle T_{t}c_{I,\sigma }\left( t\right) c_{I,\sigma }^{\dag
}\left( 0\right) \right\rangle $. The averaging of the LDOS over all sites
gives
\begin{equation}
\overline{A}\left( \omega \right) =1/N_{c}\sum_{I}A_{I}\left(
\omega \right) , \label{ADOS}
\end{equation}%
where $N_{c}$ is the size of a supercell. Also of interest is the
quasiparticle (QP) density of states
\begin{equation}
N\left( \omega \right) =\frac{1}{V_{RBZ}}\int dk\sum_{l}\delta \left( \omega
-E_{k}^{l}\right),  \label{QPDOS}
\end{equation}%
where $l$ denotes all the quasiparticle bands in the reduced
Brillouin zone (RBZ), $V_{RBZ}$ is the volume of RBZ, $k\subset
$RBZ. This latter is the density of states which determines the
sum of the quasiparticle eigenvalues which enters the ground state
energy in mean field theory. The results for these DOS in the
various modulated phases are presented below. First we consider the cases of a dSC with array of $\pi$DW and of a SDW separately and then the results when both orders coexist.

\textbf{(a) anti-phase dSC}

\begin{figure}[t]
\includegraphics
[width=15.0cm,height=15.0cm,angle=270]
{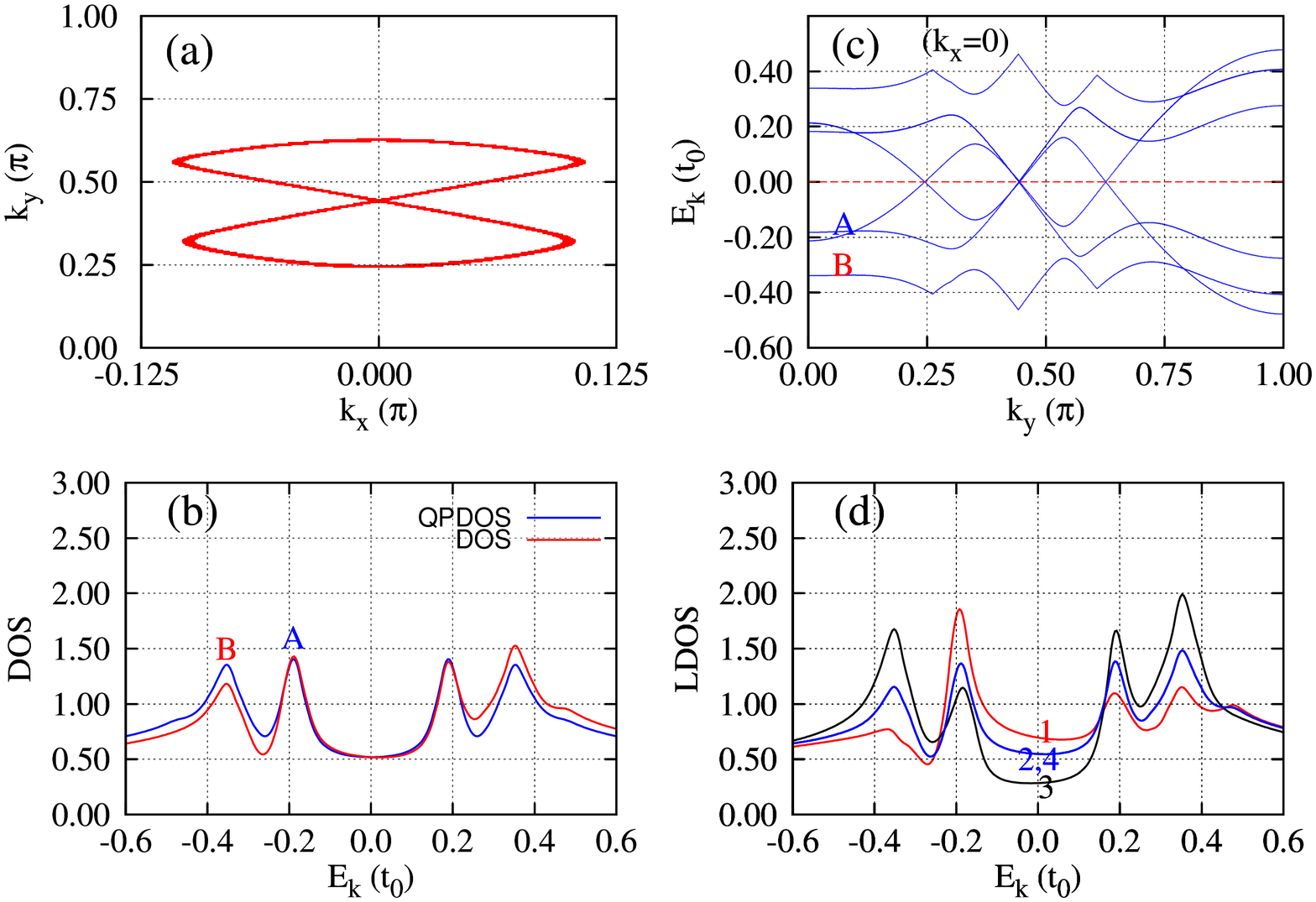}
\caption{(Color online) dSC state with site-centered anti-phase domain wall but without antiferromagnetism (doping
concentration $\protect\delta =1/8$, $t^{\prime }=0.0$, $V=0$, and a
supercell $L_{x}=8$, the energy unit is $t_{0}=300meV$). The pattern for the pairing amplitude $\Delta $ is
similar to the case shown in Fig[\protect\ref{fig:DWconfig}]. (a) Fermi
surface in the reduced Brillouin zone. (b) Quasiparticle (QP) DOS (blue) and average DOS (red). The two peaks A and B at
negative energies are a consequence of the flat dispersion along $k_{y}$
direction [shown in panel (c)] formed by the propagating of Andreev bound state along y
direction. (d) Local DOS (LDOS), near to the Fermi level the largest portion of the
density of states is at the center of the domain wall.}
\label{fig:PDRVB}
\end{figure}

We start with the DOS for an array of $\pi $DW with a superlattice
periodicity of 8 and an average hole density of 1/8. The LDOS is shown in
Fig[\ref{fig:PDRVB}], for the 3 independent lattice sites, site 1 at the $\pi $%
DW, site 3 halfway between the $\pi $DW and the remaining
equivalent sites 2, 4. In the energy region near zero, the
prominent features are a finite LDOS at all sites, which is
largest at the center of a $\pi $DW (site 1) and two sharp peaks
(labeled as A and B) symmetrically placed at positive and negative
energies. The finite LDOS at $E=0$ implies a finite quasiparticle
Fermi surface in this APdSC$^{s}$ state. The quasiparticle energy
dispersion is quite complex and is illustrated in
Fig[\ref{fig:PDRVB}c]. Along the high symmetry line, $k_{x}=0$, in
RBZ there are 3 nodal points. These expand into nodal lines for
finite $k_{x}$ to create two closed Fermi loops shown in
Fig[\ref{fig:PDRVB}a]. The two sharp peaks labeled A and B in the
DOS, $\overline{A}\left( \omega \right) $, can
be shown to originate from the almost flat bands displaced away from zero energy in Fig[%
\ref{fig:PDRVB}c]. The LDOS that appears in Fig[\ref{fig:PDRVB}d] shows
clearly an enhanced DOS near zero energy which implies a substantial energy
cost to introduce the $\pi $DW into a uniform dSC state.

\textbf{(b) SDW}

\begin{figure}[t]
\includegraphics
[width=6.0cm,height=15.0cm,angle=270]
{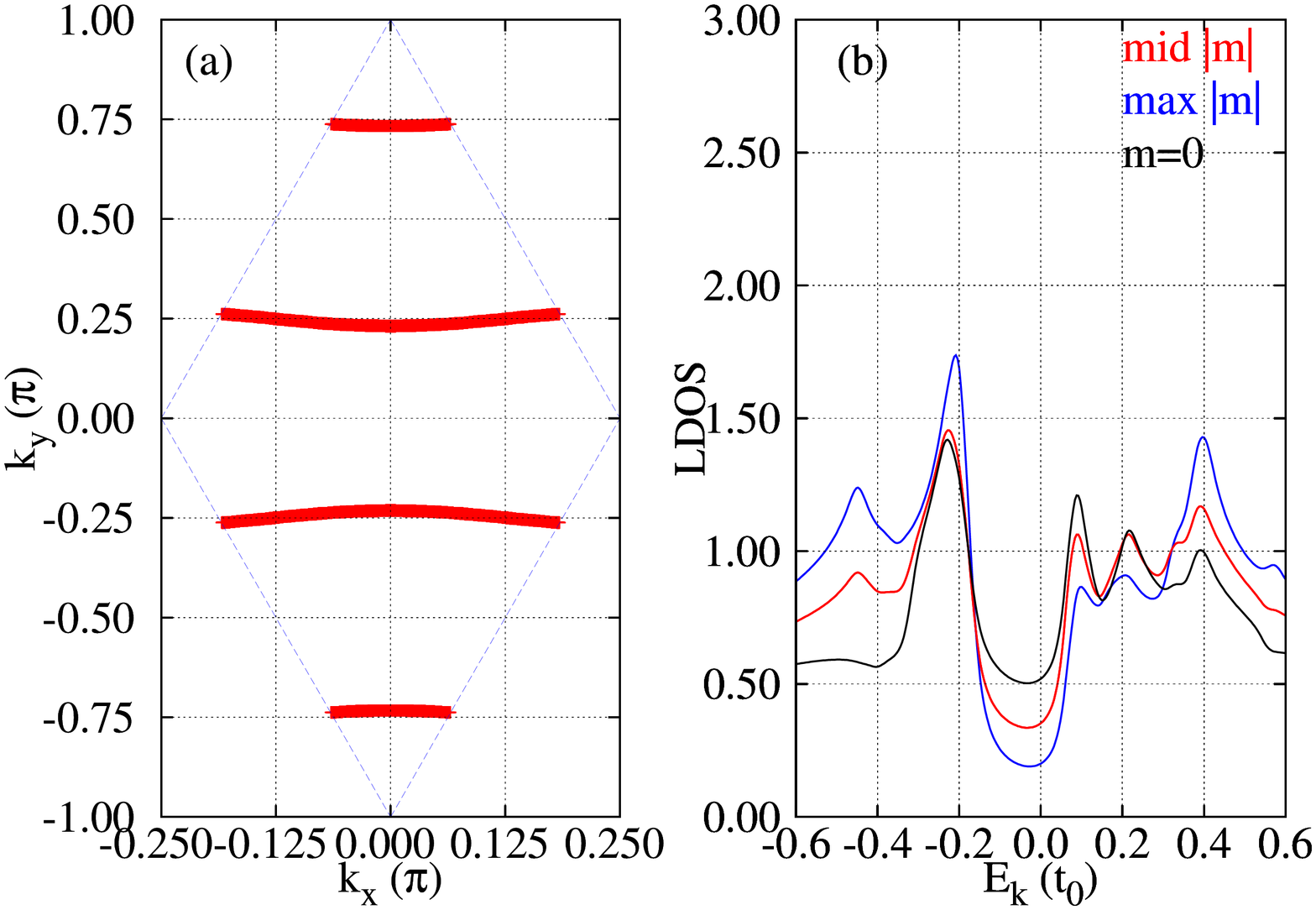}
\caption{(Color online) SDW state (without dSC $\Delta =0$) with a periodicity of $L_{x}=8$ and an average
doping concentration $\protect\delta =1/8, t^{\prime}=-0.10$ (the energy is in unit of $t_{0}=300meV$). The
antiferromagnetic sublattice moment pattern is the same as that shown in Fig[%
\protect\ref{fig:DWconfig}]. (a) Fermi surface in reduced Brillouin Zone. (b) Local density of states (LDOS). Note that
this SDW state is a metallic state.}
\label{fig:DW_noSC}
\end{figure}

The second case we considered is a simple SDW state in which an
array of AFDW
is introduced to create a 8-superlattice. Again the LDOS (see Fig[\ref%
{fig:DW_noSC}]) shows finite values at zero energy with the
largest value at the center of the AFDW ($m_{i}=0$). As a
consequence this SDW state is metallic. Note a uniform state would
also be metallic at this hole concentration of $\delta =1/8$. It
is however very relevant that the SDW superlattice does not
truncate the Fermi surface completely to give an insulating state,
since then coexistence with d-wave pairing would be disfavored.
Further any coexisting state would not be superconducting. The
Fermi surface shown in Fig[\ref{fig:DW_noSC}a] consists of
standing waves along $k_{y}$ \textit{i.e.} perpendicular to the
AFDW and two one-dimensional bands propagating along AFDW.

\textbf{(c) Coexisting SDW, CDW and dSC or anti-phase dSC}

\begin{figure}[t]
\includegraphics
[width=11.0cm,height=16.0cm,angle=270]
{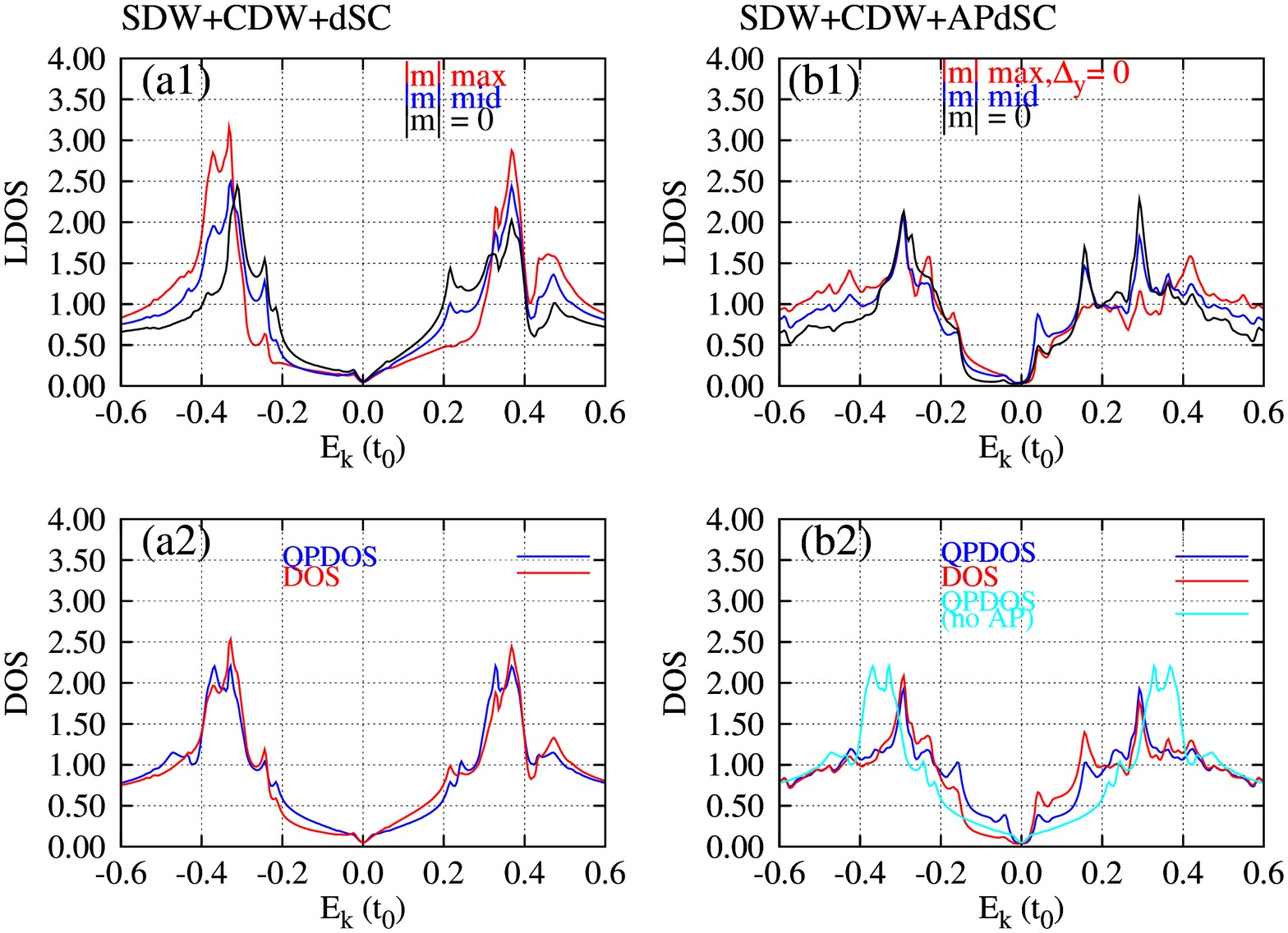}

\caption{(Color online) SDW+CDW+dSC$^{s}$ and SDW+CDW+APdSC$^{s}$ states (without or with site-centered anti-phase domain wall). The
upper figures (a1, b1) show the local density of states (LDOS) at the three inequivalent sites with
max $|m|$, zero $|m|$, and the middle site.  (doping $%
\protect\delta =1/8$, $t^{\prime }=-0.1$, $V=0$, and isotropic
$t_{x}=t_{y}$). The energy is in unit of $t_{0}=300meV$. The lower figures (a2, b2) show the average DOS and
quasiparticle (QP) DOS. In order to facilitate the
comparison between the states with and without the domain wall in
panel (b2) the cyan curve is the QP DOS for the SDW+CDW+dSC$^{s}$ state,
replotted from panel (a2). A small gap opens at zero energy.
However, a substantial part of the DOS located at lower energy is
pushed to closer to the Fermi level. This may be the reason that
the opening of a gap in the SCW+CDW+APdSC$^{s}$ state does not lead to a
lower energy relative to the state without the domain wall. Note
that a small broadening $\protect\delta =0.004t_{y}$ is used to
smooth the curve.
The nodal behavior in SDW+CDW+dSC$^{s}$ state is not a general phenomenon. For larger $%
t^{\prime }$, anisotropic $t_{x(y)}$, or external additional potential this
nodal structure may disappear. Also for other cases, (e.g. $t^{\prime
}=0,t_{x}=1,V=0$) no gap opens in SDW+CDW+APdSC$^{s}$ state.}
\label{fig:SDWCDWDRVB}
\end{figure}

We examine the coexisting state to look for possible synergy between the SDW
and dSC and also to compare the two possibilities for the superconducting
uniform dSC and the APdSC$^{s}$, \textit{i.e.} superconductivity without and with an array of $%
\pi $DW. The favorable choice of the relative position of the two domain
walls is to stagger the $\pi $DW and AFDW as shown in
Fig[\ref{fig:DWconfig}]. From Fig[\ref{fig:SDWCDWDRVB}a,b] one
sees that in both cases the LDOS develops a strong minimum around
zero energy with even a drop to zero in a narrow range around zero
energy. The site dependence of the LDOS is weaker than in the
previous cases. This strong energy minimum indicates a certain
synergy between the SDW and dSC which can lower the energy through
a truncation of the finite Fermi surface that exists in the both
cases separately, SDW and
APdSC$^{s}$. The energy difference in the LDOS between the two cases with and without $%
\pi $DW, is small. But when one compares the total ground state
energy, a finite energy cost to introduce $\pi $DW into
the superconductivity always appears.

 The strong minimum in the DOS at the Fermi level in the SDW+CDW+APdSC$^{s}$ state is consistent with the spectra obtained in  angle resolved photoemission (ARPES) and
 scanning tunnelling (STM) experiments on La$_{\text{1.875}}$ Ba$_{\text{0.125}}$CuO$_{\text{4}}$  reported by Valla \textit{et al} \cite{valla-science-06}. Our calculations give a complex quasiparticle dispersion associated with the
 8-fold superlattice which does not seem to be resolved in the ARPES spectra. So a more detailed comparison with experiment is not possible at this time but the main feature of the experimental DOS is reproduced in our calculations.

\section{discussion}

Anti-phase domain wall or $\pi$DW generally cost considerable energy in a
superconductor because they generate an Andreev bound state at the
Fermi energy due to the interference between reflected electrons
and holes. This effect is illustrated in Fig[\ref{fig:toy}a] which
shows a peak in the LDOS centered on an isolated $\pi$DW. In an
array of parallel $\pi$DW this DOS peak broadens into a
2-dimensional band due to both the propagation of the ABS along
the $\pi$DW, as illustrated in Fig[\ref{fig:toy}b], and the
overlap of the ABS on neighboring $\pi$DW. This leads to structure
which can lead to a pronounced minimum in the LDOS in certain
cases such as the case of a closely spaced array of $\pi$DW shown
in Fig[\ref{fig:PDRVB_L4}b]. This structure in the LDOS lowers the
energy cost to introduce $\pi$DW in the dSC, but leaves it still positive. For the period 8
supercell the modification of the DOS is less important. As
illustrated in Fig[\ref{fig:PDRVB}c] the APdSC$^{s}$ bandstructure is
quite complex and displays a finite Fermi surface (see
Fig[\ref{fig:PDRVB}a]). The resulting LDOS has a
finite value at the Fermi energy which is largest at the center of the $\pi$%
DW.

In the case of coexisting SDW and CDW one must first consider how the effect
of these superlattices alone. The results are presented in Fig[\ref%
{fig:DW_noSC}] which shows a metallic state with a finite DOS and Fermi
surface. This is important since if the SDW resulted in an insulating
groundstate, the addition of Cooper pairing would be less energetically
favorable and would not change the state from insulating to superconducting.
The bandstructure consists of standing waves in the direction perpendicular
to the AFDW which are propagating in the direction parallel to the AFDW.

Coexisting SDW and dSC leads to a substantial interplay between the
two broken symmetries. Recently Agterberg and Tsunetsugu showed
that there can be a synergy between the two orders due to the
presence of cross terms involving both order parameters in a
Landau expansion \cite{Agterberg-08}. The cross term depends
crucially on the relative orientation of the wavevector of the SDW
and APdSC. For the case of parallel $\mathbf{q}$-vectors under consideration
here (eg. as illustrated in Fig[\ref{fig:BDWconfig}]), however the cross
term vanishes. Nonetheless in the present case there is still a
considerable synergy between the two broken symmetries. This shows
up in the DOS as a pronounced dip at the chemical potential as illustrated in
Fig[\ref{fig:SDWCDWDRVB}b]. However, this effect is not confined
to case of APdSC but is also present in the case of a uniform
phase dSC coexisting with SDW as illustrated in
Fig[\ref{fig:SDWCDWDRVB}a]. We have not found a simple explanation
for this synergy. The quasiparticle bands in the vicinity of the
Fermi energy have a complex dispersion for which we do not have a
simple interpretation. Remarkably the form of the DOS near the
Fermi energy is very similar for coexisting SDW and dSC with and
without the array of $\pi$DW the dSC. This subtle difference in
the DOS shows up as only a small difference in the ground state
energy so that the energy cost of introducing $\pi$DW is very
small.

\section{Conclusions}

The small energy difference that we find agrees with the earlier
calculations reported by Himeda \textit{et al.} \cite{Himeda-PRL-02} for coexisting SDW and
APdSC$^{s/b}$. These authors used a VMC method in which the strong coupling onsite
constraint is exactly treated whereas here it is only approximated through
the Gutzwiller factors. This suggests that our failure to find a clear
explanation for the stabilization of APdSC$^{s/b}$ does not result from the
Gutzwiller approximation but may be because the t-J model omits some
relevant physical effect. Alternatively the special cross term between SDW
and APdSC order found by Agterberg and Tsunetsugu \cite{Agterberg-08} which favors oblique
wavevectors for the two periodicities may mean that our simple pattern with
parallel arrays of AFDW and $\pi$DW is not optimal, although on the surface
it looks very plausible to simple stagger the two domain walls. After completing this paper, we learned that a related work was posted by Chou \textit{et al.} \cite{TKLee-08}.

\section{acknowledges}
We are grateful to John Tranquada, Alexei Tsvelik and Daniel
Agterberg for stimulating discussions. KYY, TMR and MS gratefully
acknowledge financial support from the Swiss Nationalfonds through
the MANEP network. This work was also in part supported by RGC at
HKSAR (FCZ and WQC).


\begin{thebibliography}{9}
\bibitem{simplify-G} In the original form introduced by Himeda and Ogata \cite{Himeda-Ogata}, $g_{\langle i,j\rangle
}^{s,xy}=g_{i}^{s,xy}g_{j}^{s,xy}a_{\langle i,j\rangle }^{-7}$. The factor $a_{\langle i,j\rangle }$ is very complicated but takes a value very close to 1. We set
$a_{\langle i,j\rangle }=1$ for simplicity.
\end{thebibliography}
\end{document}